\documentclass[journal]{IEEEtran}
\ifCLASSOPTIONcompsoc
  \usepackage[nocompress]{cite}
\else
  \usepackage{cite}
\fi

\usepackage{graphicx}

\usepackage{amsmath}
\usepackage{comment}
\usepackage{amsthm}
\usepackage{color}
\usepackage{balance}
\usepackage{bbm}

\usepackage[normalem]{ulem}

\newtheorem{prop}{Proposition}

\usepackage{algorithm, algorithmic}

\makeatletter
\newcommand\fs@norules{\def\@fs@cfont{\bfseries}\let\@fs@capt\floatc@ruled
  \def\@fs@pre{}%
  \def\@fs@post{}%
  \def\@fs@mid{\kern3pt}%
  \let\@fs@iftopcapt\iftrue}
\makeatother
\floatstyle{norules}
\restylefloat{algorithm}
\usepackage{subfigure}

\begin{document}



\title{Performance of Offloading Strategies in Collocated Deployments of Millimeter Wave NR-U Technology}

\author{Anastasia~Daraseliya,
Eduard~Sopin, Dmitri~Moltchanov, \\Yevgeni Koucheryavy,~\IEEEmembership{Senior Member,~IEEE} and Konstantin~Samouylov\vspace{-3mm}

\thanks{A. Daraseliya, E. Sopin, and K. Samouylov are with Peoples' Friendship University of Russia (RUDN University), Moscow, Russia. Email:~{daraselia-av}@rudn.ru, {sopin-es}@rudn.ru, {samuylov-ke}@rudn.ru.}
\thanks{D. Moltchanov and Y. Koucheryavy are with Tampere University, Finland. Email:~{dmitri.moltchanov}@tuni.fi. 
}
\thanks{E. Sopin and K. Samouylov are also with Institute of Informatics Problems, Federal Research Center Computer Science and Control
of Russian Academy of Sciences, Moscow, Russia.}}

\maketitle



\begin{abstract}
5G New Radio (NR) technology operating in millimeter wave (mmWave) band is expected to be utilized in areas with high and fluctuating traffic demands such as city squares, shopping malls, etc. The latter may result in quality of service (QoS) violations. To deal with this challenge, 3GPP has recently proposed NR unlicensed (NR-U) technology that may utilize 60 GHz frequency band. In this paper, we investigate the deployment of NR-U base stations (BS) simultaneously operating in licensed and unlicensed mmWave bands in presence of competing WiGig traffic, where NR-U users may use unlicensed band as long as session rate requirements are met. To this aim, we utilize the tools of stochastic geometry, Markov chains, and queuing systems with random resource requirements to simultaneously capture NR-U/WiGig coexistence mechanism and session service dynamics in the presence of mmWave-specific channel impairments. We then proceed comparing performance of different offloading strategies by utilizing the eventual session loss probability as the main metric of interest. Our results show non-trivial behaviour of the collision probability in the unlicensed band as compared to lower frequency systems. The baseline strategy, where a session is offloaded onto unlicensed band only when there are no resources available in the licensed one, leads to the best performance. The offloading strategy, where sessions with heavier-than-average requirements are immediately directed onto unlicensed band results in just $2-5\%$ performance loss. The worst performance is observed when sessions with smaller-than-average requirements are offloaded onto unlicensed band.
\end{abstract}


\begin{IEEEkeywords}
NR-U, New Radio, overflow traffic, offloading, QoS, queuing theory, Markov chains
\end{IEEEkeywords}

\section{Introduction}


The recently standardized 5G millimeter wave (mmWave) New Radio (NR) technology promises to deliver extreme rates at the air interface \cite{parkvall2017nr}. To this aim, these systems will be utilized mainly in areas with high traffic demands, such as shopping malls, concert halls, and stadiums. In these deployment conditions traffic demands tend to fluctuate leading to quality-of-service (QoS) violations and even drop of sessions. The service process is further complicated by mmWave-inherent specifics, including the use of highly directional antennas \cite{petrov2017interference}, large propagation losses \cite{umi}, and blockage phenomenon \cite{gapeyenko2016analysis}.


To address the abovementioned challenge, a number of approaches have been suggested in the literature. Network densification using conventional NR base stations (BS) \cite{bhushan2014network} or utilizing integrated access and backhaul (IAB) architecture \cite{madapatha2020integrated} are two straightforward options. However, it leads to high capital expenditures, low resource utilization and may require additional interference mitigation techniques \cite{kovalchukov2018evaluating}. The use of mobile BSs mounted on, e.g., unmanned aerial vehicles (UAV) \cite{lagum2017strategic,gapeyenko2018flexible} or cars \cite{rabieekenari2017autonomous,petrov2019analysis} is another plausible option. However, this approach is limited to those use-cases where traffic fluctuations happen at the minutes or even tens of minutes timescale such that the network operator has sufficient time budget to densify the network. Furthermore, this option is not available in indoor environments.

We consider an alternative approach capable of smoothing small-scale traffic dynamics at seconds and sub-seconds timescales. Particularly, we investigate a joint implementation of licensed NR operating at 28 GHz and unlicensed technology occupying the spectrum at 60 GHz at a single BS. This technology, known as NR-unlicensed (NR-U), started to be addressed in 3GPP in 2019. Similarly to the LTE operation in the unlicensed band, known as LTE licensed assisted access (LAA), 3GPP has extended 5G NR to operate in the unlicensed bands. This is implemented by utilizing dynamic frequency selection (DFS) with carrier aggregation techniques and complementing then with the random access mechanism. The former two mechanisms are similar to those utilized in LTE LAA \cite{mukherjee2016licensed}. These functionalities have been further clarified in Release 16 TR 38.889 and TR 38.716 to form the 5G NR-U technology. A random-access mechanism is aimed to enable fair coexistence with IEEE 802.11ad/ay technologies. 



The performance assessment of NR-U systems is often limited to analysis of coexistence strategies or offloading functionality onto unlicensed spectrum as further discussed in Section \ref{sect:rel}. The former set of studies usually neglects system-level specifics while the latter ones often utilize simplified models for abstraction of coexistence mechanism. Furthermore, only a few models capture inherent specifics of mmWave including directional antennas utilized in both licensed and unlicensed bands as well as inherent blockage phenomenon. However, owing to principally different access mechanisms utilized in the unlicensed band, performance guarantees may not be always provided. This raises questions related to the choice of the optimal offloading strategy and density of NR-U mmWave BSs to support a given density of NR-U user equipment (UE) and UEs utilizing in unlicensed band only.


We study the service process of NR-U users in collocated mmWave NR-U/WiGig BSs with licensed and unlicensed technologies in presence of competing WiGig traffic. The assumption of having collocated NR-U BS and WiGig access points (AP) allows to target two use-cases: (i) operator own WiGig APs that can be installed at the same site as NR-U BS and (ii) third-party APs, where the considered deployment provides the worst-case interference. To this aim, we utilize the tools of stochastic geometry to characterize UE traffic demands and Markov chain theory to capture the random access dynamics in the unlicensed band by accounting for mmWave-specific propagation. Then, we employ queuing system with random resource requirements to access the rate provided to NR-U UEs as well as session drop probability in the unlicensed band under different offloading strategies. These metrics allow us to deduce the ultimate metric of interest -- required density of NR-U BSs for a given density of NR-U and WiGig UEs. The main contributions of our study are:
\begin{itemize}
    \item{an analytical framework for evaluation of the NR-U session loss probability in collocated deployments of mmWave NR-U BSs under different offloading strategies in a field of WiGig UEs;}
    \item{analysis of minimal density of collocated mmWave NR-U BSs under different session offloading strategies onto unlicensed band satisfying the prescribed session rate;}
    \item{numerical analysis demonstrating that: (i) the blockage probability inherent for mmWave frequencies heavily affects system performance (ii) the "baseline" and "fat" offloading strategies lead to almost identical performance in terms of eventual session loss probabilities.}
\end{itemize}


We structure the rest of the paper in the following way. The related work is provided in Section \ref{sect:rel}. Then, in Section \ref{sect:system} we formulate our system model. We developed our performance evaluation framework in Section \ref{sect:math}. Comparison of the considered offloading strategies is provided in Section \ref{sect:num}. Finally, conclusions are stated.


\section{Related Work}\label{sect:rel}

3GPP has introduced NR-U as a concept in 2016 in technical reports TR 38.889 and TR 38.716. Further, as the work on NR specifications progressed, NR-U standards have evolved as well. Similarly to LAA technology, the main building blocks for NR-U are carrier aggregation and dynamic frequency selection capabilities. To enable fair coexistence with technologies utilizing unlicensed bands, i.e., IEEE 802.11ad/ay, random channel access procedures have been emphasized in TR 38.889 as critical functionality.




\subsubsection{Coexistence Mechanisms}

The literature on performance analysis of NR-U technology can be divided into two directions. The first direction investigates random access coexistence mechanisms between NR-U and unlicensed technologies. There are two approaches proposed so far to address this challenge: (i) duty-cycle based approaches and (ii) listen-before-talk (LBT) based mechanisms. The study in \cite{patriciello2020nr} explores LBT mechanism originally proposed in TR 38.889 in 3GPP Release 15 and reveal that it may not fully fulfill the requirements on coexistence between NR-U and IEEE 802.11ad WiGig technology under realistic NR implementation. The authors in \cite{lagen2018listen} proposed a new version of coexistence mechanism, called listen-before-receive (LBR), and proceeded to compare it to that defined in 3GPP TR 38.889. In spite of this mechanism allows to eliminate performance degradation issues of the original LBT coexistence scheme, the fairness was found to be non-perfect. Another extension has been suggested in \cite{ali2018fair}, where a channel observation LBT (CoLBT) random access mechanism has been proposed. The main aim of the CoLBT scheme was to improve fairness between NR-U and unlicensed technologies.

Further extension of the LBT scheme for NR-U technology is due to Zhang et al \cite{zhang2020hybrid}, who proposed a hybrid scheme, where conventional LBT from TR 38.889 is utilized in congested channel conditions while no special access method is utilized when the shared unlicensed channel is relatively free. The carried our simulations have shown that this approach potentially provides much smaller latency and throughput. Performance optimization of the carrier sensing threshold for conventional LBT from TR 38.889 is performed in \cite{oni2020optimized}, where the authors reported closed-form results. Finally, excellent overview and comparison of various NR-U channel access proposals are provided in \cite{lagen2019new}.

In our paper, we accept the following definition of fairness "for $N$ UEs utilizing the unlicensed band, the resources delivered to WiGig UEs should not be different in both cases: (i) $N-M$ of these UEs are native WiGig UEs, (ii) $N-M$ of these UEs are NR-U UEs."

\subsubsection{Resource Allocation and User Performance}

The second research direction in context of NR-U that received much less attention so far is related to the issues of resource allocation and user performance in NR-U environment. Among others, the authors in \cite{lu20195g} considered the application of NR-U technology to extended virtual reality (VR) applications at sporting events. Their results illustrate that provided fair coexistent mechanism system-level performance improvements in user-centric and operator-centric metrics can be achieved. An analytical performance evaluation model for the system in \cite{lu20195g} has been developed in \cite{lu2019integrated}. Among other conclusions, the authors revealed that  proportional splitting of traffic between the licensed and unlicensed mmWave bands leads to better performance. However, in their study, the authors considered only a single NR-U BS and also did not include the specifics of random access scheme in the unlicensed band in their model. In spite of focusing on LAA technology, the authors in \cite{markova2019performance,maule2018delivering} developed and analyzed a simple duty-cycle based scheme for fair resource splitting between LTE-U and Wi-Fi operating in the unlicensed band.


The question of fair coexistence in unlicensed bands has been considered in \cite{bairagi2018game,bairagi2018qoe}. The unique feature of the study is that the proposed solution accounts for QoS metrics of applications. Specifically, the authors formulated an optimization problem such that QoS is maintained while unlicensed users are shielded from LTE-U users and solved it using the game-theoretic and Q-learning approaches. Technically, the proposed solution can be implemented by utilizing the duty-cycle coexistence approach.


In our recent study \cite{daraceliya2020pimrc}, we developed the model for collocated NR-U design explicitly capturing the random access behavior in unlicensed band and characterizing the NR-U session loss probability. However, the scope of the interest has been limited to a single offloading strategy, where NR-U session is only offloaded onto unlicensed band only when there is an insufficient amount of resources in the licensed band. This strategy may not be optimal. Furthermore, the authors utilized a simple M/M/1/K queuing model to capture the specifics of resource allocation in the licensed band. 

\subsubsection{Summary}

Summarizing, we note that those studies concentrating on coexistence mechanisms do not address user traffic dynamics and issues related to user performance in NR-U systems. Contrarily, the authors investigating resource allocation aspects and offloading strategies generally assume perfect coexistence mechanisms or resort to time-consuming system-level simulations. In this paper, we fill this gap by proposing a mathematically tractable approach capable of capturing both aspects in detail. Specifically, we relax simplifying assumptions in \cite{daraceliya2020pimrc} by considering accurate model of the service process in the licensed band and evaluating performance of different offloading strategies.

\section{System model}\label{sect:system}

We now proceed to introduce the system model. We start with the deployment model and description of the coexistence mechanism and then proceed with mmWave-specific propagation, blockage, and antenna models. Traffic and association models are introduced next. Finally, we define the considered offloading strategies and metrics of interest. Notation used in our study is provided in Table~\ref{tab:notation}.


\begin{table}[!t]
\caption{Notation used in the paper.}
\label{tab:notation}
\begin{center}
\begin{tabular}{p{0.19\columnwidth}p{0.71\columnwidth}}
\hline
\textbf{Parameter}            	& \textbf{Definition}   \\
\hline\hline
$\chi_{B,N},\chi_{B,W}$ & Density of pedestrians \\
\hline
$\lambda$ & Arrival intensity from the Poisson process\\
\hline
$\lambda_A$,$\lambda_B$ & Densities of NR-U BSs /  pedestrians\\
\hline
$h_A$ & Heights of NR-U BS \\
\hline
$h_U,h_B$ & Heights of UE and blockers \\
\hline
$T$ & Number of retransmission attempts\\
\hline
$S(x)$ & SINR at UE\\
\hline
$L(x),L_{dB}(x)$ & Path loss in linear and dB forms\\
\hline
$P_T$ & Emitted power\\
\hline
$G_{N,A},G_{W,A}$ & Array gains at NR-U/WiGig BSs\\
\hline
$G_{N,U},G_{W,U}$ & Array gains at NR-U/WiGig UEs\\
\hline
$M_I$ & Interference margin\\
\hline
$f_c$ & Operational frequency\\
\hline
$A_i,\zeta_i$ & Constants related to propagation models\\
\hline
$\alpha$ & Antenna array's HPBW\\
\hline
$N_{\cdot}$ & Number of antenna elements in appropriate plane\\
\hline
$\phi_{3db},\phi_m$ & Antenna array parameters\\
\hline
$S_{min}$ & Outage threshold\\
\hline
$S_{B}(x),S_{nB}(x)$ & SINR at distance $r$ in blocked/non-blocked states\\
\hline
$r_B$ & Average radius of human blockers\\
\hline
$N_0$ & Thermal noise\\
\hline
$M_{S,B},M_{S,nB}$ & Fading margins in blocked/non-blocked states\\
\hline
$\sigma_{S,B}$ & Standard deviation of shadow fading in blocked state\\
\hline
$\text{erfc}^{-1}(\cdot)$ & Inverse complementary error function\\
\hline
$p_b(r),p_b$ & Blockage probabilities\\
\hline
$p_c$ & Collision probability\\
\hline
$r_N,r_W$ & Coverages of NR-U BS and WiGig AP\\
\hline
$B(x)$ & CDF of service time \\
\hline
$E[S_e]$ & Mean spectral efficiency can be obtained as follows\\
\hline
$\alpha$ &  Half-power beamwidth of antenna array \\
\hline
$\beta$ & The array orientation (azimuth angle)\\
\hline
$\theta_{3db}$ & $\pm{}3$ dB point of antenna array\\
\hline
$\theta_m$ & Location of the array maximum\\
\hline
$N$ & Number of antenna elements \\
\hline
$R$ & Number of resource units \\
\hline
$R_F$, $R_S$ & Offloading thresholds for the "fat/slim" strategy\\
\hline
$K$ & Maximum number of sessions in the system \\
\hline
$\pi_W$   & Transmission probability of WiGig UE \\
\hline
$\pi_N$ & Transmission probability of NR-U UE\\
\hline
$\Pi_N$ & Successful probability of transmission \\
\hline
$\pi_{\star}$ & Probability that the second type session is rerouted \\
\hline
$\pi_{\star}$ &  Probability of session loss at NR-U\\
\hline
$b_i$ & Amount of slots in state $i$\\
\hline
$b_{\min}$ & Amount of resources required by NR-U\\
\hline
$R_{\min}$ & Minimum requested session rate\\
\hline
$W$ & Contention window size\\
\hline
$p_{l,j}$ & pmf of required resources by session of type $l$\\
\hline
$p_{j,\star}$ & pmf of required resources in the unlicensed band\\
\hline
$p_{j,\star}$ & pmf of required resources for strategy $\star$ \\
\hline
$p_{j,\star}$ & pmf of required resources in the licensed band\\
\hline
$P_{k}(r)$ & Stationary state probability\\
\hline
$q_i$ & State $i$ stationary probability\\
\hline
$\theta$ &  Successful  transmission probability\\
\hline
$m_j$ & Spectral efficiency \\
\hline
$R_{sU,j}^{N}$ & Rate achieved by the session  \\
\hline
$E[R_{sU}^N]$  & Attained rate in the unlicensed band of NR-U UE\\
\hline
$Q_{sU}$ & NR-U UE session loss probability for strategy $s$\\
\hline
$Q_s$ & Eventual session loss probability\\
\hline
\end{tabular}
\end{center}
\vspace{-7mm}
\end{table}

\subsection{Deployment Model}\label{Sys_mod_DM}

\begin{figure}[t!]
\vspace{-0mm}
\centering
\includegraphics[page=1,clip, trim=0.9cm 8.3cm 2.1cm 0.1cm, width=1.0\columnwidth]{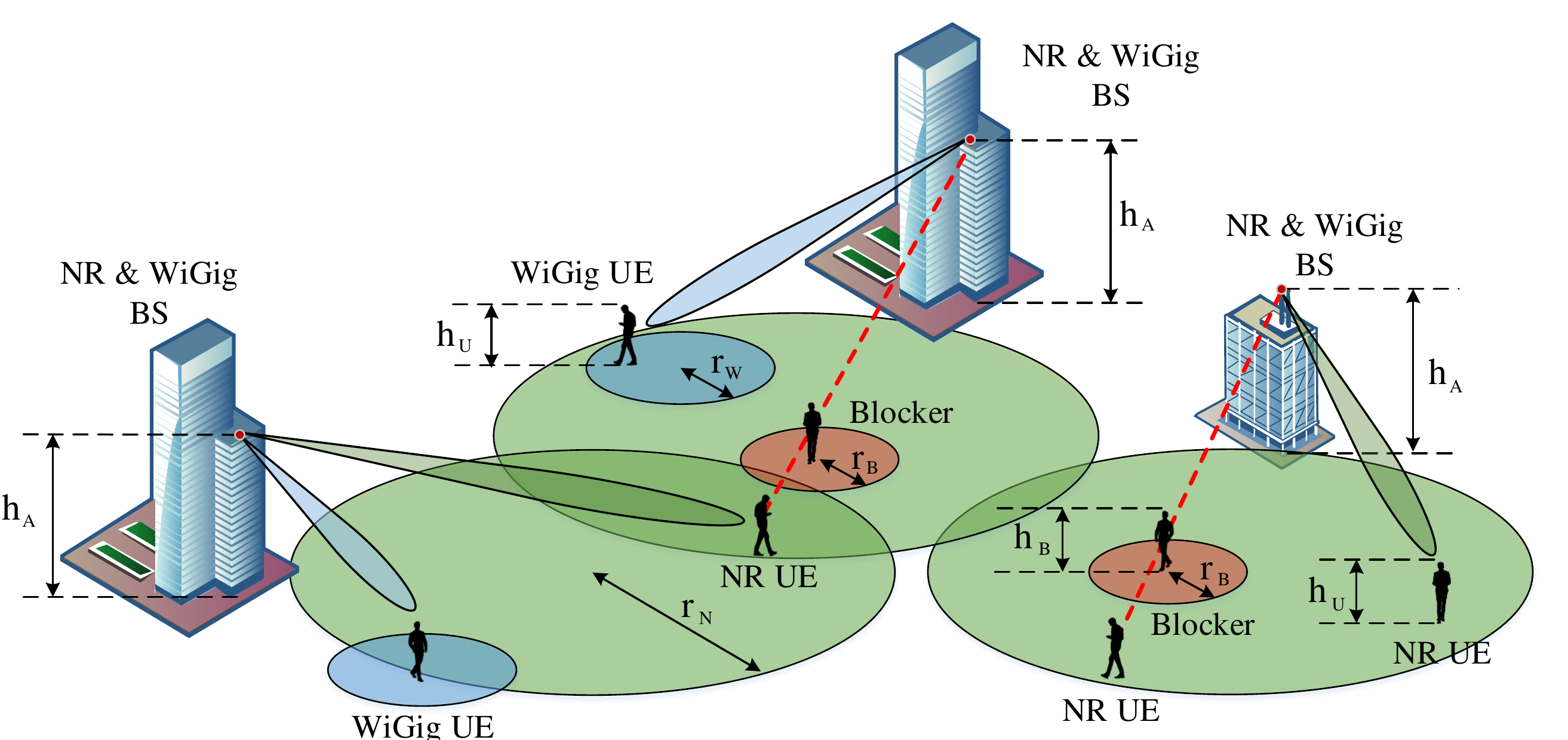}
\caption{The considered deployment model.}
\label{fig:environment}
\vspace{-4mm}
\end{figure}

Consider deployment with NR-U BS and WiGig AP collocated on the same site, see Fig. \ref{fig:environment}. The licensed part of NR-U BS uses the licensed 28 GHz band with the bandwidth $B_N$ MHz \cite{3gppband}. The unlicensed part of NR-U BS operates in 60 GHz band using one channel with the bandwidth of $B_W$ MHz \cite{WiGigband}. Each site also contains a WiGig AP operating in the same channel. In what follows, we refer to these collocated sites as NR-U BS implicitly implying presence of WiGig APs at each of those. NR-U BSs follow Poisson point process in $\Re^2$ with the density of $\lambda_A$. NR-U BS height is $h_B$. 


We consider two types of UEs, NR-U UEs and WiGig UEs. Both types of UEs follow mutually independent PPPs in $\Re^2$ with densities $\chi_{B,N}$ and $\chi_{B,W}$. NR-U UEs may utilize both unlicensed and licensed bands while WiGig UEs may only utilize the former band. The rules of accessing the bands are specified below in Section \ref{sect:offloadStrat}. Note that we assume no internal cooperation between NR-U and WiGig at the site.


\subsection{NR-U Coexistence Mechanism}

\begin{figure}[t!]
\vspace{-0mm}
\centering
\includegraphics[page=1,clip, trim=0cm 15.3cm 19.9cm 0cm, width=1.0\columnwidth]{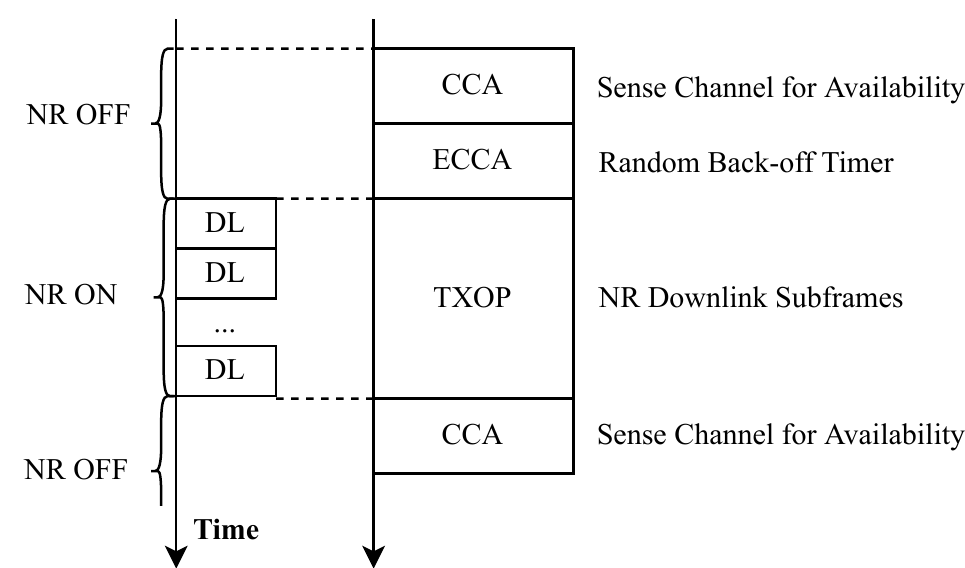}
\vspace{-4mm}
\caption{CoLBT coexistence mechanism for NR-U.}
\label{fig:coLBT}
\vspace{-2mm}
\end{figure}

We assume that NR-U UEs utilize the improved version of the LBT coexistence mechanism from 3GPP 38.889 specification -- CoLBT defined in \cite{ali2018fair}, see Fig. \ref{fig:coLBT}. According to CoLBT, NR-U UE that has a packet ready for transmission first chooses a back-off counter uniformly in $(0,CW)$, where $CW$ is the current length of the contention window (CW). CW is decremented at each slot. Once it reaches $0$, packet transmission is initiated. Note that CW counter is paused if the medium is sensed to be busy. 

The result of the transmission can be successful or unsuccessful. The latter may happen as a result of (i) collision with another NR-U or WiGig UE transmission or (ii) LoS blockage. Note that the collision may only happen when another NR-U or WiGig UE transmits from the same sector currently covered by WiGig AP antenna. In both cases, CW is doubled. The number of retransmission attempts is upper bounded by $T$. At each unsuccessful retransmission attempt, CW is doubled. When the maximum number of retransmission attempts $T$ is reached, packet is considered to be lost.





\subsection{Blockage, Propagation, and Antenna Models}

\subsubsection{Blockage Model}



We assume that the propagation path to NR-U BS or WiGig AP can be blocked by human bodies. Humans move in $\Re^2$ by following random direction mobility (RDM, \cite{nain2005properties}) model with mean run time $\tau$ and speed $v$. Humans are represented by cylinders with height $h_B$ and radius $r_B$. The UE height is assumed to be $h_U$, $h_U<h_B$. In what follows, we will utilize the line-of-sight (LoS) blockage probability, $p_b(r)$, in the following form \cite{gapeyenko2016analysis}
\begin{equation}\label{eqn:blockPropDist}
p_b(r)=1-e^{-2\lambda_B r_B\left(r\frac{h_B-h_U}{h_{(\cdot)}-h_U}+r_B\right)},
\end{equation}
where $h_{(\cdot)}$ is NR-U BS or WiGig AP height while $\lambda_B$ is the human density.

Note that we assume that blockage always leads to erroneous reception in WiGig. First of all, even the minimum reported human body blockage attenuation of 15 dB is a large 
value requiring drastic change in modulation and coding scheme (MCS) \cite{slezak2018empirical}. The second reason is that the frequency of beamtracking is lower in WLANs as compared to NR systems, and thus, the signal drop may not be detected timely. Still, it is feasible to remove this effect assuming that the blockage probability is zero or at least smaller than the actual blockage probability. The latter would allow to represent non-perfect beamtracking.








\subsubsection{Propagation Model}




The signal-to-interference-plus-noise ratio (SINR) at the NR-U or WiGig UE is written as
\begin{align}\label{eqn:genericProp}
S(y)=\frac{P_{N,U}G_{N,A}G_{N,U}}{L(y)M_S(N_0W+M_I)},
\end{align}
where $y$ is the separation distance, $G_{N,A}$ and $G_{N,U}$ are the transmit and receive gains, $N_0$ is the thermal noise, $M_I$ and $M_S$ are the interference and shadow fading margins, $L(y)$ is the path loss at the distance $y$, and $N_0$ is the thermal noise. Note that instead of utilizing interference margin $M_I$ one may utilize detailed complex models such as \cite{kovalchukov2018analyzing,petrov2017interference}. Also, note that shadow fading margin differ in blocked $M_{S,B}$ and non-blocked states and $M_{S,nB}$, see \cite{umi}.





Since we consider mmWave licensed and unlicensed bands, we utilize the same 3GPP urban-micro (UMi) street-canyon propagation model for both bands \cite{umi}, i.e.,
\begin{align}\label{eqn:plDb}
\hspace{-2mm}L_{dB}(y)=
\begin{cases}
32.4 + 21.0\log(y) + 20\log{f_{M,c}},\,\text{non-bl.},\\
32.4 + 31.9\log(y) + 20\log{f_{M,c}},\,\text{blocked},\\
\end{cases}\hspace{-2mm}
\end{align}
where $f_{M,c}$ and $y$ are the carrier frequency and separation distance between communicating entities. In what follows, we represent the path loss as $A_iy^{-\zeta_i}$, $i=1,2$, where $A_1=A_2=10^{2\log_{10}{f_{M,c}}+3.24}$ $\zeta_1=2.1$, $\zeta_2=3.19$.


By accounting for blockage in (\ref{eqn:blockPropDist}), SINR is now given by
\begin{align}\label{eqn:prop}
S(y)&=\frac{P_{N,U}G_{N,A}G_{N,U}}{(N_0 W+M_I)A}\Big[\frac{y^{-\zeta_1}}{M_{S,nB}}[1-p_b(y)]\nonumber\\
&+\frac{y^{-\zeta_2}}{M_{S,B}}p_b(y)\Big],
\end{align}
and using $C=P_{N,U}G_{N,A}G_{N,U}/((N_0 W+M_I) A)$, we have
\begin{align}\label{eqn:prop_final}
S(y)=\frac{Cy^{-\zeta_1}}{M_{S,nB}}(1-p_{b}(y))+\frac{C y^{-\zeta_2}}{M_{S,B}}p_{b}(y).
\end{align}


\subsubsection{Antenna Model}

By following \cite{kovalchukov2018evaluating}, for both licensed and unlicensed bands we consider the cone antenna model, where the half-power beamwidth (HPBW) of the main lobe is represented by conical zone. The HPBW of the main lobe, $\alpha$, is given by $\alpha = 2|\theta_m - \theta_{3db}|$ \cite{constantine2005antenna}, where $\theta_{3db}$ is the $3$ dB point, $\theta_m$ is the array orientation. For horizontal orientation, $\theta_m=\pi/2$. Note that $\alpha$ in horizontal and vertical planes can be closely approximated \cite{constantine2005antenna} by $102^{\circ}/N_{\cdot}$, where $N_{\cdot}$ is the number of antenna elements in the appropriate plane.




The gain of antenna array is given by \cite{constantine2005antenna}
\begin{align}\label{eqn:G}
G = \frac{1}{\theta_{3db}^+-\theta_{3db}^{-}}\int_{\theta_{3db}^-}^{\theta_{3db}^+} \frac{\sin(N\pi\cos(\theta)/2)}{\sin(\pi\cos(\theta)/2)}d\theta.
\end{align}





\subsection{Traffic, Associations, and Offloading Schemes}\label{sect:offloadStrat}

For NR-U UEs we assume that the rate required from the network is $R_{\min}$. This corresponds to rate-greedy non-adaptive applications such as video streaming, AR/VR, remote telepresence. Depending on the distance from NR-U UE to NR-U BS maintaining the rate $R_{\min}$ may require different amount of physical resources, see Section \ref{sect:math}. The intensity of NR-U sessions is $\lambda$ sessions per second. WiGig traffic is assumed to be fully elastic, that is, the rate can be dynamically adapted to changing conditions.  


Both NR-U UEs and WiGig UEs are associated with collocated NR-U BSs and WiGig APs based on the average reference signal receive power (RSRP). In practice, it implies that physically nearest site is selected.  






We consider the following offloading schemes:
\begin{itemize}
    \item{\textit{Baseline.} In this case, NR-U UEs try to associate with the nearest NR-U BS and utilize licensed band. If there is no sufficient amount of resources to accept the session for service, it is rerouted to the unlicensed band. If the current rate provided to the offloaded session in this band is less than $R_{\min}$ the session is dropped.}
    \item{\textit{"Fat" offloading.} In this case, the decision upon the service point of NR-U UE is taken upon arrival, based on the amount of resources needed to satisfy the minimum rate requirements $R_{\min}$. If these resources are higher than a certain value $E[R_{\min}]+r_T$, where $r_T$ is some threshold, then the session is served in the unlicensed band. Otherwise, the session first attempts to reserve resources at the licensed band of NR-U BS and only if there is insufficient amount of resources it is offloaded to the unlicensed band.}
    \item{\textit{"Slim" offloading.} This scheme is a mirrored version of the previous one, where NR-U UE upon arrival is routed to the unlicensed band of NR-U BS if the amount of resources needed to achieve the minimum rate $R_{\min}$ is smaller than $E[R_{\min}]-r_T$, where $r_T$ is some threshold. Otherwise, the session is first attempted at the licensed band of NR-U BS and only if there are not enough resources there routed to the unlicensed band.}
\end{itemize}

The rationale behind the latter two schemes is that the achieved rate of sessions in the unlicensed band is a non-linear function of the number of competing UEs. Note that only those sessions that are closer to the NR-U BS than $r_W$, where $r_W$ is the coverage of unlicensed technology, can be offloaded to the unlicensed band.


\subsection{Metrics of Interest}

The main metric of interest in our study is the so-called eventual NR-U session loss probability. This metric is defined as the probability to drop a session due to inability to maintain the minimum rate $R_{\min}$ at NR-U BS utilizing both licensed and unlicensed bands. Note that this metric is calculated differently for considered offloading strategies. By utilizing this metric we report on the density of NR-U BSs required to maintain a certain eventual NR-U session loss probability in presence of competing WiGig traffic.


\section{Performance Evaluation Framework}\label{sect:math}


In this section, we develop and solve the model for considered offloading strategies. We start with the overview of the modeling approach. Then, we characterize the resource request distribution of NR-U sessions to licensed part of NR-U BS, and define the service processes at the licensed and unlicensed parts of NR-U BS. Using these models we finally evaluate the NR-U UE eventual session loss probability for considered offloading strategies.

\begin{figure*}[!t]
\vspace{-0mm}
\footnotesize
\begin{align}\label{eqn:sfsnr}
W_{{S_{SF}}}(x) &= \frac{1}{{2r_N^2}} \Big[ {{A^{2/\zeta }}{{10}^{ - \frac{x}{{5\zeta}}}}{e^{\frac{{{\sigma ^2}{{\log }^2}(10)}}{{50{\zeta^2}}}}}}\Big[ {{\rm{erf}}\left( {\frac{{50\zeta \log A - 25{\zeta^2}\log B + {\sigma ^2}{{\log }^2}10 - 5\zeta x\log 10}}{{5\sqrt 2 \zeta \sigma \log (10)}}} \right) - } \nonumber\\
	&{ - {\rm{erf}}\left( {\frac{{50\zeta (\log A - \zeta \log ({h_B} - {h_U})) + \sigma _S^2{{\log }^2}10 - 5\zeta x\log (10)}}{{5\sqrt 2 \zeta \sigma \log (10)}}} \right)}\Big]+ \left( {r_N^2 + {{({h_B} - {h_U})}^2}} \right) {\rm{erf}}\left( {\frac{{ - 10\log A + 5\zeta \log B + x\log 10}}{{\sqrt 2 \sigma \log (10)}}} \right)-\nonumber\\
	&{ - {{({h_B} - {h_U})}^2}{\rm{erf}}\left( {\frac{{\sqrt 2 ( - 10\log A + 10\zeta \log ({h_B} - {h_U}) + x\log 10)}}{{\sigma \log 100}}} \right) + r_N^2} \Big].
\end{align}
\normalsize
\hrulefill
\vspace{-0mm}
\end{figure*}

\subsection{Resource Request Characterization}\label{sect:resources}


To characterize the session service processes in the licensed and unlicensed bands we need to determine the probability that a NR-U session can be offloaded onto unlicensed band and the amount of requested resources by NR-U and WiGig UE sessions in both bands.




\subsubsection{Coverage} 

In order to derive the abovementioned quantities, we first need to determine the coverage area of NR-U BS in licensed and unlicensed bands, $r_N$ and $r_W$, respectively. Consider $r_N$ first and observe that it is can be bounded by both intersite distance, $r_{N,V}$, and SINR, $r_{N,S}$, that is, $r_{N}=\min(r_{N,S},r_{N,V})$. The latter can be determined by utilizing the propagation model in the worst possible conditions, i.e., blocked state. Specifically, SINR threshold in the LoS blocked state, $S_{th}$ at the distance $r_{N,S}$ corresponding to the worst MCS \cite{nrmcs} takes the form
\begin{align}\label{eqn:dist_00}
S_{th}= C\left(r_{N,S}^2+(h_B-h_U)^2\right)^{-\frac{\zeta }{2}}.
\end{align}

Solving for $r_{N,S}$, we obtain the sought radius
\begin{align}\label{eqn:11}
r_{N,S} = \sqrt{\left(\left(S_{th}M_{S,B}\right)/C\right)^{\frac{2}{\zeta}}-(h_B-h_U)^2}.
\end{align}
where the shadow fading margin in the $M_{S,B}$ can be found as $M_{S,B}=\sqrt{2}\sigma_{S,B}\text{erfc}^{-1}(2p_C)$. Here, $\sigma_{S,B}$ is the shadow fading standard deviation in the blocked state \cite{umi}, $p_C$ is the fraction of time in outage at the edge of the cell. 





The coverage in the unlicensed band, $r_{W,S}$ is estimated similarly. The second component in $r_{N}$, the coverage induced by intersite distance, $r_{N,V}$ is dictated by the Voronoi cells in $\Re^2$ organized by the NR-U BS deployment process in $\Re^2$. We utilize circular approximation to the Voronoi cells to estimate $r_{N,V}$. As the closed-form solution for the area of a Voronoi cell is not available we rely upon approximations in \cite{tanemura2003statistical}.





\subsubsection{Resource Request Distribution}

We now proceed characterizing with the resource request distribution by NR-U UE. The procedure involves derivation of the SINR distribution and further discretization by NR MCSs \cite{nrmcs}. We start with deriving SINR cumulative distribution function (CDF). To this aim, we first differentiate between LoS blocked and non-blocked state deriving SINR CDFs for them and then weight the obtained distributions with blockage probability.

The probability density function (pdf) of two-dimensional distance from NR-U UE to NR-U BS is $w_{R}(x)=2x/2r_N$ \cite{moltchanov2012distance}. Now by utilizing random variables functional transformation technique \cite{ross2014introduction}, cumulative distribution function (CDF) of three-dimensional distance, $D$, is provided by
\begin{align}\label{eqn:3ddistance}
W_{D}(x)=\frac{{2{h_B}{h_U} -h_B^2+ h_U^2 + {x^2}}}{r_N^2},
\end{align}
defined for ${h_B} - {h_U}< x < \sqrt {r_N^2 + h_B^2 - 2{h_B}{h_U} + h_U^2}$.

The SINR in decibel scale can be found using the same technique by utilizing transformation in the form $\phi_{S,dB}(x)=10 \log _{10}\left(Ax^{-\zeta}\right)$, where $X$ is three-dimensional distance obtained in (\ref{eqn:3ddistance}). We arrive at 
\begin{align}\label{eqn:deriv}
W_{S^{dB}}(x)=1 - \frac{{{{10}^{ - \frac{x}{{5\zeta }}}}{A^{\frac{2}{\zeta}}} - {{\left( {{h_B} - {h_U}} \right)}^2}}}{{r_N^2}},
\end{align}
defined for all $x>10\log_{10}[{A{(r_N^2 + {{( {{h_B} - {h_U}})}^2})^{-\zeta/2}}}]$.

Recall that the shadow fading is characterized by Normal distribution \cite{rappaport1996wireless}. Thus, the SINR in decibel scale accounting for shadow fading impairments can be written as $S_{SF} = {S^{dB}} + \mathcal{N}(0,\sigma)$. Thus the sought SINR pdf can be determined by utilizing convolution operation as follows
\begin{align}
	{W_{{S_{SF}}}}(y) = \int_{ - \infty }^\infty  {{W_S}\left( {y + u} \right)} \frac{{{e^{ - \frac{{{u^2}}}{{2{\sigma ^2}}}}}}}{{\sqrt {2\pi } \sigma }}du.
\end{align}

The latter can be represented in terms of error function, $\text{erf}(\cdot)$, as in (\ref{eqn:sfsnr}), where $B={r_N^2 + {{\left( {{h_B} - {h_U}} \right)}^2}}$. To obtain SINR CDFs corresponding to LoS blocked and non-blocked states, $W_{S_{nB}}$ and $W_{S_{B}}$, one shall utilize different standard deviations of shadow fading, $\sigma_{S,B}$ and $\sigma_{S,nB}$, and different path loss exponents, $\zeta_1$ and $\zeta_2$.

Having $W_{S_{nB}}$ and $W_{S_{B}}$ at our disposal allows us to determine the overall SINR CDF by weighting two branches with the blockage probability, i.e.,
\begin{align}\label{eqn:fullsnr}
W_S(x)=p_bW_{S_{B}(x)}+\left(1-p_b\right)W_{S_{nB}(x)}.
\end{align}

Now, the spectral efficiency of NR-U session in the licensed band can be determined by partitioning of the SINR CDF according to NR MCS \cite{nrmcs} the target block error rate (BLER). Multiplying it by the minimum rate $R_{\min}$ we obtain the session resource requirements. Also, observe that only those sessions that are located closer than $r_{W}$ can be potentially offloaded to the unlicensed band. The resource requirements pmf of these sessions can be obtained according to the abovementioned procedure by replacing $r_{N}$ by $r_{W}$. Finally, the pmf of resource requirements of a session in the unlicensed band can be determined similarly by utilizing MCSs to SINR mapping defined for IEEE 802.11ad/ay technologies. Alternatively, the mean spectral efficiency is obtained by
\begin{align}\label{eqn:spectralEffWiGig}
E[S_e]=\int_{0}^{r_N}\frac{2x}{r_N}\log_{2}(1+S(y))dx,
\end{align}
where $S(y)$ is SINR in the unlicensed band. Note that WiGig UEs experience similar spectral efficiency. By utilizing the minimum requested rate $R_{\min}$ we can further determine the mean amount of resources requested by NR-U session in the unlicensed band as $b_{\min}=R_{\min}/E[S_e]$.

\begin{figure}[b!]
\vspace{-3mm}
\centering
\includegraphics[page=1,clip, trim=0cm 24.6cm 10.1cm 0.11cm,width=1.0\columnwidth]{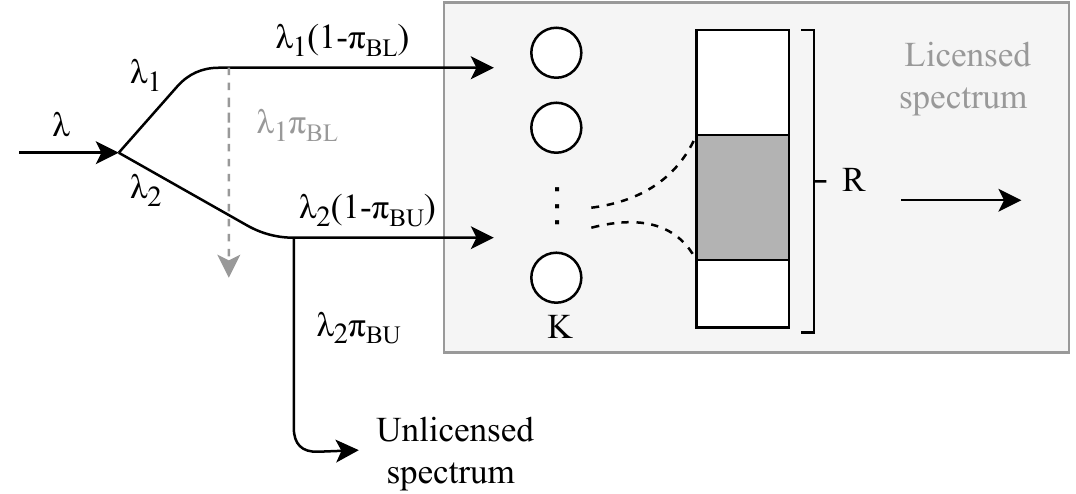}
\caption{Illustration of the queuing model for the "baseline" strategy.}
\label{fig:queue5}
\vspace{-0mm}
\end{figure} 

\subsection{Offloading Schemes Specifics and Arrival Rates}


Following our methodology, we can analyze all three defined offloading strategies using the same framework consisting of two principal components: (i) queuing system with random resource requirements, and (ii) random access analysis of the unlicensed shared medium. For the "baseline" strategy, see Fig. \ref{fig:queue5}, the loss probability at NR-U BS is, in fact, the offloading probability to the unlicensed band, while the eventual loss probability is the probability that the requested minimum rate is not satisfied there. For the latter two strategies, "fat" and "slim" offloading, see Fig. \ref{fig:queue5_Fat}, the decision about the first service point, licensed or unlicensed band, is taken upon arrival of NR-U UE session. Thus, for these two schemes, the eventual loss probability is defined as a probability that the minimum rate $R_{\min}$ is not satisfied in the unlicensed band. 


Accounting for inherent imbalance between coverage radii of licensed and unlicensed bands, $r_{N}$ and $r_{W}$ (i.e., $r_N>r_W$) we now introduce two types of sessions. The first type of sessions can be served in the licensed band only. When there are not enough resources to service this session, the session is lost. Geometrical locations of these sessions are in the ring $r_W<x<r_N$. Sessions of the second type can be potentially offloaded to the unlicensed band as they originate from the circle with radius $r_W$ around the NR-U BS. 


Recalling the notation introduced in Section \ref{sect:system}, let $\lambda=\chi_{B,N}r_{N}^2\pi p_N\lambda_S$ be the initial intensity of NR-U sessions, where $\chi_{B,N}$ is the density of NR UEs per square meter defined in Section \ref{Sys_mod_DM}, $r_N^2\pi$ is  the BS coverage area for the licensed technology, and $\lambda_S$ is the intensity of sessions from NR-U UE. Since $r_W\leq{}r_N$, we represent $\lambda$ as $\lambda=\lambda_1+\lambda_2$, where $\lambda_1$ and $\lambda_2$ are the arrival rates of NR-U sessions from $0<x<r_W$ circle and $r_W<x<r_W$ ring, respectively. From now on, they are referred to as first and second types of sessions.

\subsubsection{Baseline Strategy}

For the baseline strategy, the overflow arrival rate to the unlicensed band is simply the sum of the intensities of all both types of sessions, i.e., $\lambda_{BL}=\lambda_1+\lambda_2$. The fraction of load that the licensed band of NR-U BS cannot handle coincides with the overflow arrival rate of the second type of sessions to the unlicensed part of NR-U BS. These parameters are calculated as $\lambda_{BU}=\lambda_2 \pi_{BU}$, where $\pi_{BU}$ is the probability that the second type of session is routed to the unlicensed band. We also denote $\pi_{BL}$ the probability that the first type session is lost upon arrival. These probabilities are derived in the next section.

\subsubsection{"Fat" Offloading Strategy}

\begin{figure}[t!]
\vspace{-0mm}
\centering
\includegraphics[page=1,clip, trim=0cm 23.9cm 7.5cm 0cm, width=1.0\columnwidth]{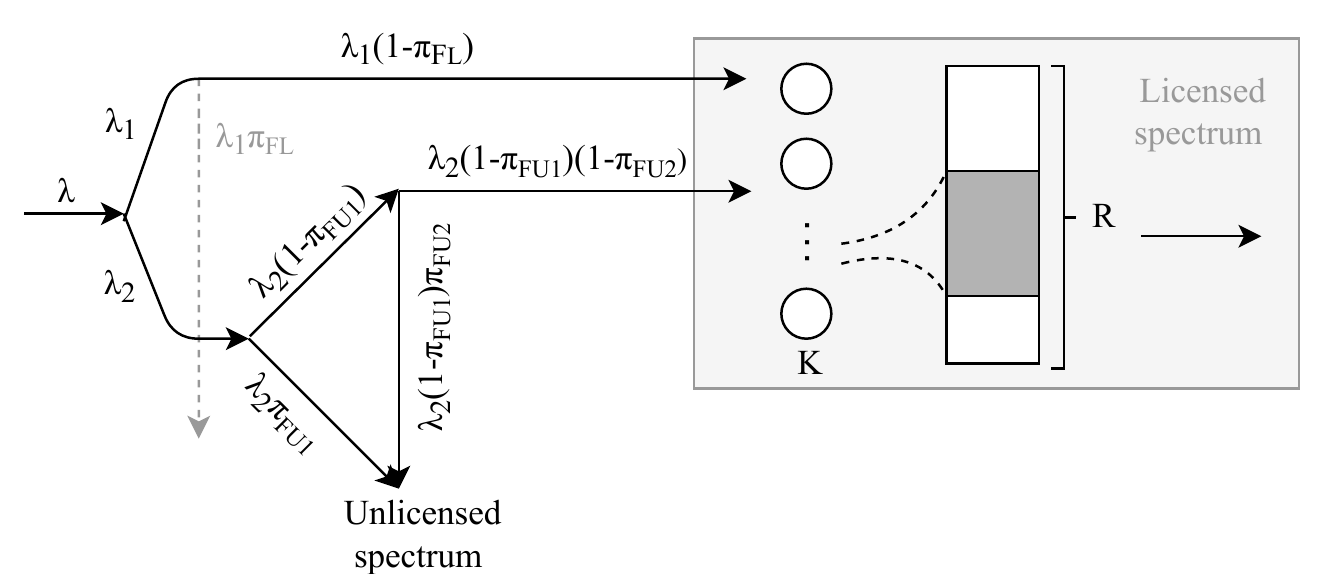}
\caption{Illustration of the queuing model for the "fat" strategy.}
\label{fig:queue5_Fat}
\vspace{-3mm}
\end{figure}

For "fat" and "slim" offloading strategies, see Fig. \ref{fig:queue5_Fat}, the pmfs of resource requirements at NR-U BS depend on the thresholds, $R_F$ and $R_S$. Particularly, for the "fat" strategy, the arrival flow of the second type of sessions is divided according to the "weight" of the session. "Heavier" sessions are initially routed to the unlicensed band with a probability $\pi_{FU1}$, and with complementary probability, $(1-\pi_{FU1})$, "lighter" sessions are routed to licensed band. Thus, the overall rate, $\lambda_{FL}$, of both types of sessions to the licensed band is $\lambda_{FL}=\lambda_1+\lambda_2\left(1-\pi_{FU1}\right)$.


Observe, that the second type sessions arrive to the unlicensed band in two cases: (i) when the "weight" of the session is more than a certain threshold $R_F$, (ii) when there are no sufficient amount of resources available for a session that has been initially routed to licensed band. The probability $\pi_{FU}$ that the second type session will be directed to the unlicensed spectrum is the sum of the probability $\pi_{FU1}$ that the session was "heavy" and the probability $\pi_{FU2}$ that a "light" session cannot be handled at the licensed band and thus offloaded to unlicensed one, i.e., $\pi_{FU}=\pi_{FU1}+(1-\pi_{FU1})\pi_{FU2}$. The probabilities $\pi_{FU1}$ and $\pi_{FU2}$ 
will be defined in the next section. Note that the arrival rate $\lambda_{FU}$ to the unlicensed part of NR-U BS can be written as $\lambda_{FU}=\lambda_2\pi_{FU}$.

\subsubsection{"Slim" Offloading Strategy}

Similarly to the "fat" strategy, for the "slim" strategy, the arrival rate $\lambda_{SL}$ to licensed  part of NR-U BS can be calculated by $\lambda_{SL}=\lambda_1+\lambda_2\left(1-\pi_{SU1}\right)$, where the probability $\pi_{SU1}$ that the session is "light", that is, session which requires less than $R_S$ amount of resources, and was originally directed to the unlicensed band.

In contrast to the "fat" strategy, sessions arrive at the unlicensed range in two cases: (i) when the "weight" of the session is less than the threshold $R_S$, or (ii) when the session has been initially routed to the licensed band but the amount of available resources is insufficient to serve it. With these in mind, the probability that the second type session will be directed to the unlicensed band is given by $\pi_{SU}=\pi_{SU1}+(1-\pi_{SU1})\pi_{SU2}$, where $\pi_{SU2}$ is the probability that a "heavy" session cannot be served in the licensed band and has to be offloaded to the unlicensed band. The probabilities $\pi_{SU1}$ and $\pi_{SU2}$ will be defined in the next section. The overall arrival rate $\lambda_{SU}$ to the unlicensed band and can be calculated as $\lambda_{SU}=\lambda_2\pi_{SU}$.

\subsection{Service Process in the Licensed Band}


To model the session service process in the licensed band, we utilize the framework of resources queuing systems \cite{sopin2015aggregation,sopinM2M2018,begishev2019quantifying}. To this aim, consider a multi-server queuing system with $K<\infty$ servers and $R<\infty$ resource units, where $K$ accounts for the limit on the number of NR-U UE sessions in the system. Sessions of two types arrive to the system, both according to the Poisson processes with arrival rates $\lambda_1$ for the first type and $\lambda_2$ for the second one. Thus, the total arriving flow is Poisson with parameter $\lambda=\lambda_1+\lambda_2$. The service time distribution is exponential with the rate $\mu$.


Service process of each session requires a server and a random amount of resources, $0 \leq r \leq R$. The distributions of resource requirements for considered session types are given by $\{ p_{l,j}\}_{j\geq 0}$, $l=1,2$, where $p_{l,j}$ is the probability that a session of type $l$ requires $j$ resources. According to \cite{sopin2015aggregation} resource-based queuing system with two flows can be analyzed as a system with one aggregated flow assuming the following
\begin{equation}\label{eqn:prob_pBL}
p_{j,BL}=\frac{\rho_1}{\rho_B}p_{1,j}+ \frac{\rho_2}{\rho_B}p_{2,j}.
\end{equation}
where the offered traffic load is $\rho_B=\rho_1+\rho_2$, $\rho_i=\lambda_i/\mu$.


The system operates as follows. An arriving session is accepted to the system if at the moment of arrival there are sufficient amount of resources available. Alternatively, an arriving session is dropped. In this case, the first type of session is lost while the session of the second type is being redirected to the unlicensed band. When the service time of a session is over, it leaves the system releasing all the occupied resources. The system behavior can be described by a stochastic process $X(t)=(\xi(t), \gamma(t))$, where $\xi(t)$ is the number of sessions in the system and $\gamma(t) =(\gamma_1(t), \gamma_2(t),\dots, \gamma_{\xi(t)}(t))$, $\gamma_i(t)$ is the vector of the amount of resources allocated to the $i$th session at the time instant $t$.


Denote by $P_{k}(r)$ the stationary probability that there are $k$ sessions in the system that totally occupy $r$ resources, i.e.,
\begin{equation}\label{eqn:syseq_prob_k_r_lim}
P_{k}(j)=\lim_{t \rightarrow \infty} P\{ \xi(t)=k, \sum_{i=1}^{\xi(t)} \gamma_i(t)=j\}, \, 0\leq j\leq R.
\end{equation}

According to \cite{NaumovRes2016}, the stationary distribution is given by
\begin{align}
\label{eqn:syseq_prob_normalization_constant}
&P_{k}(r)=P_{0}\frac{\rho_B^i}{k!}p_{r,BL}^{(k)},\,k=1,2,\dots,K,\nonumber\\
&P_{0}=\left(1+\sum_{k=1}^K\frac{\rho_B^k}{k!}\sum_{r=0}^Rp_{r,BL}^{(k)}\right)^{-1},
\end{align}
where $\{ p^{(k)}_{r,BL}\}_{r \geq 0}$ is $k$-fold convolution of pmf $\{p_{r,BL}\}_{r \geq 0}$, and is calculated as follows
\begin{equation}\label{eqn:k_fold_convolution}
p_{r,BL}^{(k)}=\sum_{j=0}^{r}p_{r-j,BL}^{(k-1)}p_{j,BL}, \text{ } k \geq 2, 
\vspace{-1mm}
\end{equation}
where $p_{r,BL}^{(1)}=p_{r,BL}$, $r \geq 0$.


The loss probability of the second type sessions, i.e., the probability that session will be rerouted to the unlicensed part of the NR-U BS, is then given by
\begin{equation}\label{eqn:block_prob_mgn0_pB}
\pi_{BU}=1-P_{0}\sum_{k=0}^{K-1}\frac{\rho_B^k}{k!}\sum_{r=0}^R \sum_{j=0}^{r} p_{2,j}
p_{r-j,BL}^{(k+1)}.
\end{equation}


For large values of $K$ and $R$ calculations according to (\ref{eqn:block_prob_mgn0_pB}) are computationally demanding. In this case, a recurrent computational algorithm developed in \cite{sopinM2M2018} can be adopted. According to it, the loss probability can be calculated as
\begin{equation}\label{eqn:block_prob_mgn0_pB_G}
\pi_{BU}=1-G_B^{-1}(K,R)\sum_{i=0}^{R}p_{2,i}G_B(K-1,R-i),
\end{equation}
where the unknown term $G_B(K,R)$ is given by
\begin{equation}
G_B(n,r)=\sum_{i=0}^{n}\frac{\rho_B^i}{i!}\sum_{j=0}^{r}p_{j,BL}^{(i)},\,P_{0}=G_B^{-1}(K,R).
\end{equation}


By utilizing the abovementioned results, the probability that there is no sufficient amount of resources in the licensed band to serve a session of the first type is calculated as follows
\begin{equation}\label{eqn:block_prob_mgn0_pBL_G2}
\pi_{BL}=1-G_B^{-1}(K,R)\sum_{i=0}^{R}p_{1,i}G_B(K-1,R-i).
\end{equation}

The described methodology for estimating the NR-U session loss probability in the licensed band remains intact for all the considered strategies. However, the choice of strategy affects input parameters including the intensity of arriving flow $\lambda$, and the distribution of the amount of requested resourced delivered in Section \ref{sect:resources} as detailed below.


For the "fat" strategy, the probability that the second type session requires $j$ resources in the licensed band is given by
\begin{equation}\label{eqn:block_prob_fat_L2}
p_{j,FL2}=\left(\sum_{i=0}^{R_{F}}p_{2,i}\right)^{-1}p_{2,j}, \text{ }0\leq j\leq R_F.
\end{equation}

Hence, similar to (\ref{eqn:prob_pBL}), the resource request distribution of sessions in the aggregated flow is provided by
\begin{equation}\label{eqn:block_prob_fat_L}
p_{j,FL}=\frac{\rho_{F1}}{\rho_F}p_{1,j}+ \frac{\rho_{F2}}{\rho_F}p_{j,FL2},
\end{equation}
where the offered traffic load $\rho_F=\rho_{F1}+\rho_{F2}$, and $\rho_{F1}=\rho_1$, $\rho_{F2}=\lambda_{2}(1-\pi_{FU1})/\mu$.

By analogy to the baseline strategy (\ref{eqn:block_prob_mgn0_pB_G}), the probability that a session of the first type cannot be served in the licensed band is calculated as
\begin{equation}\label{eqn:block_prob_mgn0_pFL_G2}
\pi_{FL}=1-G_F^{-1}(K,R)\sum_{i=0}^{R}p_{1,i}G_F(K-1,R-i).
\end{equation}


Similarly, for the "slim" strategy the probability that 
the second type session requires $j$ resources in the licensed band is given by
\begin{equation}\label{eqn:block_prob_slim_L2}
p_{j,SL2}=\left(1-\sum_{i=0}^{R_{S}}p_{2,i}\right)^{-1}p_{2,j}, \text{ } j\geq R_S.
\end{equation}

Similarly to (\ref{eqn:block_prob_fat_L}), the resource request distribution of sessions in the aggregated flow is provided by
\begin{equation}\label{eqn:block_prob_slim_L}
p_{j,SL}=\frac{\rho_{S1}}{\rho_S}p_{1,j}+ \frac{\rho_{S2}}{\rho_S}p_{j,SL2},
\end{equation}
where the offered traffic load 
$\rho_S=\rho_{S1}+\rho_{S2}$ and $\rho_{S1}=\rho_1$, $\rho_{S2}=\lambda_{2}(1-\pi_{SU1})/\mu$.

Finally, similar to the baseline strategy (\ref{eqn:block_prob_mgn0_pBL_G2}), the probability that a session of the first type cannot be served in the licensed band is calculated as follows
\begin{equation}\label{eqn:block_prob_mgn0_pSL_G2}
\pi_{SL}=1-G_S^{-1}(K,R)\sum_{i=0}^{R}p_{1,i}G_S(K-1,R-i).
\end{equation}

\subsection{Unlicensed Band Characterization}

Here, for each considered strategy, we will characterize the probability distribution of resource requirements and the intensities of session arrivals to the unlicensed band.

\subsubsection{Baseline strategy}

For the baseline strategy the probability that a session, originally lost in the licensed band of NR-U BS, requires $j$ resources in the unlicensed part, is provided by
\begin{equation}\label{eqn:block_prob_baseline}
p_{j,BU}=\frac{p_{2,j}}{\pi_{BU}}\left(\sum_{r=0}^{R}P_{K}(r)+\sum_{k=0}^{K-1}\sum_{r=R-j+1}^R P_{k}(r)\right),
\end{equation}
where $0\leq j\leq R$. 

By using $G_B(k,r)$ we can simplify (\ref{eqn:block_prob_baseline}) as follows
\begin{align}\label{eqn:block_prob_baseline_1_prof}
&G_B(K-1,R)-G_B(K-1,R)= \nonumber\\ 
&=\sum_{i=0}^{K}\frac{\rho_B^i}{i!}\sum_{j=0}^{R}p_j^{(i)}-\sum_{i=0}^{K-1}\frac{\rho_B^i}{i!}\sum_{j=0}^{R}p_j^{(i)}= 
\frac{\rho_B^K}{K!}\sum_{j=0}^{R}p_j^{(K)}.
\end{align}

Multiplying (\ref{eqn:block_prob_baseline_1_prof}) by $P_{0}$ we obtain $\sum_{r=0}^{R}P_{K}(r)$. Then, the first sum in (\ref{eqn:block_prob_baseline}) can be written as
\begin{equation}\label{eqn:block_prob_baseline_1}
\sum_{r=0}^{R}P_{K}(r)=\frac{G_B(K,R)-G_B(K-1,R)}{G_B(K,R)},
\end{equation}
while the second sum in (\ref{eqn:block_prob_baseline}) can be represented as
\begin{equation}\label{eqn:block_prob_baseline_2_prof}
\sum_{k=0}^{K-1}\sum_{r=R-j+1}^R\hspace{-2mm}P_{k}(r) =\sum_{k=0}^{K-1}\left(\sum_{r=0}^R P_k(r)-\sum_{r=}^{R-j} P_k(r)\right).
\end{equation}

Noticing that
\begin{equation}\label{eqn:block_prob_baseline_2}
\sum_{k=0}^{K-1}\sum_{r=R-j+1}^R\hspace{-3mm}P_{k}(r)=\frac{G_B(K\hspace{-1mm}-\hspace{-1mm}1,R)-G_B(K\hspace{-1mm}-\hspace{-1mm}1,R\hspace{-1mm}-\hspace{-1mm}j)}{G_B(K,R)},
\end{equation}
and further substituting (\ref{eqn:block_prob_baseline_1}) and (\ref{eqn:block_prob_baseline_2}) into (\ref{eqn:block_prob_baseline}), we see that the following proposition holds true.

\begin{prop}The probability (\ref{eqn:block_prob_baseline}) that the session offloaded to the unlicensed band requires $j$ resources for the "baseline" strategy is given by
\begin{equation}\label{eqn:block_prob_baseline_Gnr}
p_{j,BU}=\frac{1}{\pi_{BU}}p_{2,j}\frac{G_B(K,R)-G_B(K-1,R-j)}{G_B(K,R)}.
\end{equation}
\end{prop}

\subsubsection{Fat strategy}

For the "fat" strategy, the probability $\pi_{FU1}$ that the session is "heavy", i.e., session requires more than $R_F$ resources, and is thus originally routed to the unlicensed spectrum is given by
\begin{equation}\label{eqn:pi_fat_U1}
\pi_{FU1}=1-\sum_{i=0}^{R_{F}}p_{2,i}.
\end{equation}

By analogy with the loss probability in the "baseline" strategy (\ref{eqn:block_prob_mgn0_pB}), the probability $\pi_{FU2}$ that a "light" session cannot be served in the licensed band and thus offloaded to the unlicensed band can be calculated as
\begin{equation}\label{eqn:pi_fat_U2}
\pi_{FU2}=1-P_{0,F}\sum_{k=0}^{K-1}\frac{\rho_F^k}{k!}\sum_{r=0}^R \sum_{j=0}^{r}
p_{r-j,FL}^{(k+1)} p_{j,FL2} .
\end{equation}

Introducing $G_F(n,r)=\sum_{i=0}^{n}\frac{\rho_F^i}{i!}\sum_{j=0}^{r}p_{j,FL}^{(i)}$, we see that, by analogy to 
(\ref{eqn:block_prob_mgn0_pB_G}), the probability $\pi_{FU2}$ (\ref{eqn:pi_fat_U2}) is given by
\begin{equation}\label{eqn:block_prob_mgn0_pB_G_fat}
\pi_{FU2}=1-G_{F}^{-1}(K,R)\sum_{i=0}^{R}p_{i,FL2}G_F(K-1,R-i).
\end{equation}

The probability that the session requires $j$ resources in the unlicensed band needs to be calculated separately for two cases: (i) when a session is "heavy" and thus initially routed to the unlicensed band, and (ii) when a session is first routed to the licensed band but there are not enough of resources available for its service. Reflecting on these cases we arrive at
\begin{align}\label{pi_fat_U}
&\frac{1-\pi_{FU1}}{\pi_{FU}}p_{2,j}\hspace{-1mm}\left(\displaystyle\sum_{r=0}^{R}P_{K,F}(r)+\sum_{k=0}^{K-1}\sum_{r=R-j+1}^R\hspace{-4mm} P_{k,F}(r)\hspace{-1mm}\right)\hspace{-1mm},j\leq R_F,\nonumber\\
&\frac{1}{\pi_{FU}}p_{2,j},j>R_F.
\end{align}

Combining (\ref{eqn:block_prob_baseline_1}) and (\ref{eqn:block_prob_baseline_2}) with (\ref{pi_fat_U}), we see that the following proposition holds true.

\begin{prop} The probability that the session requires $j$ resources in the unlicensed band for the "fat" strategy is
\begin{equation}\label{pi_fat_U_Gkr}
\hspace{-1mm}p_{j,FU}=
\begin{cases}
\frac{1-\pi_{FU1}}{\pi_{FU}}p_{2,j}\frac{G_F(K,R)-G_F(K-1,R-j)}{G_F(K,R)},j\leq R_F,\\
\frac{1}{\pi_{FU}}p_{2,j},j>R_F.\\
\end{cases}\hspace{-10mm}
\end{equation}
\end{prop}

\subsubsection{Slim strategy}

For the "slim" strategy, the probability $\pi_{SU1}$ that the session is "light", i.e., session requires less than $R_S$ resources, and thus is initially routed to the unlicensed band is given by
\begin{equation}\label{eqn:pi_slim_U1}
\pi_{SU1}=\sum_{i=0}^{R_{S}}p_{2,i}.
\end{equation}

Similarly to the fat strategy, the probability $\pi_{SU2}$ that a "heavy" session cannot be served in the licensed band and thus offloaded to the unlicensed one can be calculated as
\begin{equation}\label{eqn:pi_slim_U2}
\pi_{SU2}=1-P_{0,S}\sum_{k=0}^{K-1}\frac{\rho_S^k}{k!}\sum_{r=0}^R \sum_{j=0}^{r} 
p_{r,SL}^{(k+1)} p_{j,SL2} .
\end{equation}

Denoting $G_S(n,r)=\sum_{i=0}^{n}\frac{\rho_S^i}{i!}\sum_{j=0}^{r}p_{j,SL}^{(i)}$, we see that by analogy to 
(\ref{eqn:block_prob_mgn0_pB_G}), the probability $\pi_{FU2}$ (\ref{eqn:pi_slim_U2}) can be written as
\begin{equation}\label{eqn:block_prob_mgn0_pB_G_slim}
\pi_{SU2}=1-G_{S}^{-1}(K,R)\sum_{i=0}^{R}p_{i,SL2}G_S(K-1,R-i).
\end{equation}

Now, similarly to (\ref{pi_fat_U}), we have for $p_{j,SU}$
\begin{align}\label{pi_slim_U}
&\frac{1-\pi_{SU1}}{\pi_{SU}}p_{2,j}\left(\displaystyle\sum_{r=0}^{R}P_K^S(r)+\sum_{k=0}^{K-1}\sum_{r=R-j+1}^R\hspace{-4mm} P_k^S(r)\right)\hspace{-1mm},j\geq R_S,\nonumber\\
&\frac{1}{\pi_{SU}}p_{2,j},\,j<R_S.
\end{align}

Finally, combining (\ref{eqn:block_prob_baseline_1}) and (\ref{eqn:block_prob_baseline_2}) with (\ref{pi_slim_U}), we establish the following proposition.

\begin{prop}The probability that the session requires $j$ resources in the unlicensed band for the "slim" strategy is
\begin{equation}\label{pi_slim_U_Gkr}
\hspace{-1mm}p_{j,SU}=
\begin{cases}
\frac{1-\pi_{SU1}}{\pi_{SU}}p_{2,j}\frac{G_S(K,R)-G_S(K-1,R-j)}{G_S(K,R)},\,j\geq R_S,\\
\frac{1}{\pi_{SU}}p_{2,j},\,j<R_S.\\
\end{cases}\hspace{-10mm}
\end{equation}
\end{prop}

\subsection{Service Process in the Unlicensed Spectrum}

\begin{figure}[t!]
\vspace{-0mm}
\centering
\includegraphics[width=1.0\columnwidth]{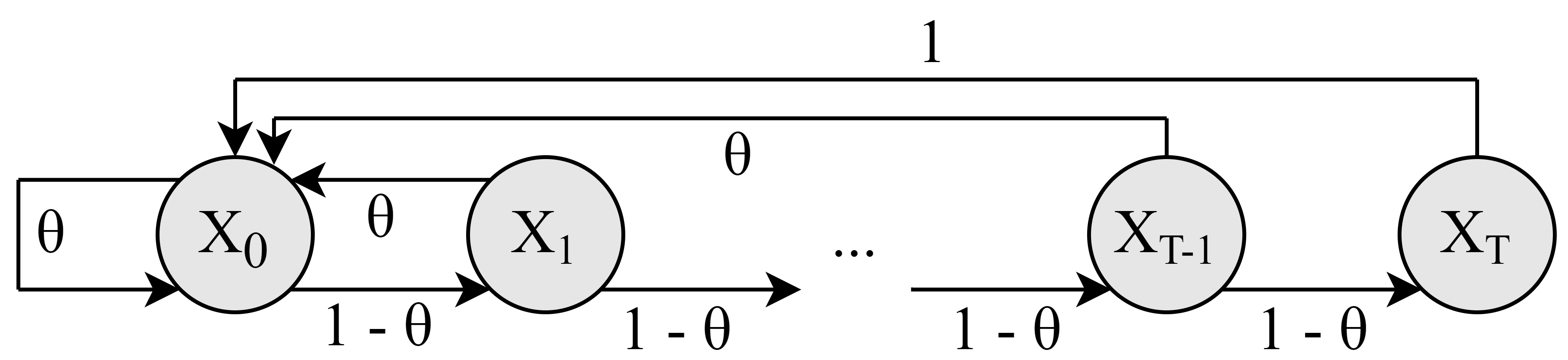}
\caption{State transition diagram of the Markov model.}
\label{fig:queue1}
\vspace{-4mm}
\end{figure} 

Recall that for all considered strategies, NR-U session can be offloaded to the unlicensed band, where it competes with WiGig sessions for transmission resources. To obtain the probability that for offloaded NR-U session the minimum rate guarantees $R_{\min}$ are satisfied, we need to first determine the successful packet transmission probability for NR-U and WiGig UEs in the unlicensed band. Note that in this section, we first define the main inter-dependencies between auxiliary probabilities for a fixed number of competing NR-U ($n_N$) and WiGig ($n_W$) UEs. Then, the auxiliary probabilities are evaluated for all possible values of $n_N$ and $n_W$ to obtain the successful transmission probability.


The probability $\theta=\theta(n_N,n_W)$ that a transmission attempt is successful is written as
\begin{align}\label{eqn:Successful_transmission_prob}
\theta=(1-p_{c})(1-p_b),
\end{align}
where $p_c=p_c(n_N,n_W)$ and $p_b$ are collision and LoS blockage probabilities.


We capture the competition dynamics between NR-U and WiGig UEs in the unlicensed band by utilizing the Markov process $\{X_n, n=0,1,\dots,T\}$, where $T$ is the maximum number of retransmission attempts. The state of the process $X_j=j$ corresponds to the transmission attempt. The transition diagram of the process is shown in Fig. \ref{fig:queue1}. By introducing  $q_i=\lim_{n\to\infty} P\{X_n=i\},\,n=0,1,\dots,T$ one can write the balance equations for this process in the following form
\begin{align} \label{eqn:lemma1_proof1}
\hspace{-3mm}\begin{cases}
q_{0} =q_{0} \theta +\dots+q_{T-1} \theta+q_{T},\\
q_{i} =q_{i-1} (1-\theta ), \text{  } i=1,2,\dots,T-1,\\
q_{T} =q_{T-1} (1-\theta ).\\
\end{cases}
\end{align} 

The system (\ref{eqn:lemma1_proof1}) has the closed-form solution given by
\begin{align}\label{eqn:lemma1_proof5_q1}
&q_{i} =\frac{(1-\theta )^{i}\theta}{1-(1-\theta )^{T+1} },\, i=0,1,\dots,T.
\end{align}

Now, we introduce the probabilities that WiGig UE and NR-U UE transmit in a randomly chosen slot -- $\pi_W=\pi_W(n_N,n_W)$ and $\pi_{N}=\pi_N(n_N,n_W)$. For a fixed number of NR-U and WiGig UEs, $n_{N}$ and $n_{W}$, respectively, the collision probability is given by
\begin{equation} \label{eqn:pc} 
p_{c} =1-(1-\pi _{W})^{n_{W}}(1-\pi _{N})^{n_{N}}.
\end{equation} 

We now determine two unknown terms in (\ref{eqn:pc}) -- $\pi_{N}$ and $\pi_{W}$. To determine these quantities, observe that UEs transmit in states $X_j=j$ only. Thus, $\pi_N$ can be found by determining the ratio between the average slot duration and the overall amount of time slots spent by UEs, that is,
\begin{align}\label{eqn:lemma2_proof2}
    \pi_{N} &=\Bigg[
    \sum_{i=0}^{T}q_j b_j
    \Bigg]^{-1},
\end{align}
where $b_j$ is the average amount of slots in state $j$, i.e.,
\begin{align} \label{eqn:b_j}
b_{j} =\sum _{i=1}^{2^{j} W}\frac{1}{2^{j} W} i =\frac{2^{j} W+1}{2}, j = 0,1,..,T.
\end{align}

By utilizing (\ref{eqn:lemma1_proof5_q1}),  (\ref{eqn:b_j}) into (\ref{eqn:lemma2_proof2}) and simplifying the NR-U transmission probability $\pi_N$ can be found as
\begin{align}\label{eqn:lemma2_proof2_2}
\pi_{N} &=\Bigg[\frac{1}{2}+
\frac{\left(1-2^{T+1} \left(1-\theta \right)^{T+1} \right)\theta W}{2(2\theta -1)(1-(1-\theta )^{T+1})}
\Bigg]^{-1}.
\end{align}

By solving the set of nonlinear equations \eqref{eqn:Successful_transmission_prob}, \eqref{eqn:pc} and \eqref{eqn:lemma2_proof2_2} with different parameters $n_N$ and $n_W$, we can obtain the corresponding probabilities for all possible numbers of active NR-U and WiGig UEs.

Let us now denote by indices $B$, $F$, and $S$ the type of the considered strategy, "baseline", "fat", and "slim", respectively. By utilizing NR-U UE transmission probability $\pi_N$ we can obtain the successful transmission probability as
\begin{align}\label{eq:U_W}
\Pi_{N,s}&= 
\sum_{i=0}^{\infty}
\frac{\left(\rho_{N,s}^{\star}\right)^i}{i!}e^{-\rho_{N}^{\star}} 
\sum_{j=0}^{\infty}
\frac{(\rho_{W}^{\star})^j}{j!}\times{}\nonumber\\
&\times{}e^{-\rho_{W}^{\star}}
\pi_{N}(i,j)\theta(i,j),
\end{align}
where $\rho_{N,s}^{\star}=\lambda_{sU}/\mu$ and $\rho_W^{\star}=\lambda_{W}/\mu_W$ are the total offered load on the licensed and unlicensed bands. 

Now, the rate attained by NR-U UE in the unlicensed band is
\textcolor{black}{
\begin{align}\label{eq:RjFU}
R_{sU,j}^{N}=\Pi_{N,s}B_U m_j,
\end{align}}
where $m_j$ is the spectral efficiency, while the mean value is
\begin{align}\label{eq:Rate_E_R}
E[R_{sU}^N]=\sum_{j=0}^{R}p_{j,sU}\Pi_{N,s}B_U m_j.
\end{align}

\subsection{Eventual NR-U Session Loss Probability}


Having obtained the mean rate of NR-U UEs in the unlicensed band, we can determine the eventual NR-U session loss probability. To this aim, we define $Q_{sU}$ to be NR-U UE session loss probability for strategy $s$. This probability provides indication that the minimum rate $R_{\min}$ is not met in the unlicensed band. By using $R_{sU,j}^{N}$, (\ref{eq:Rate_E_R}) becomes limited by rate threshold $R_{\min}$ and the sought metric is given by
\begin{align}\label{eq:QF}
Q_{sU}=P\{ R<R_{min}\}=\sum_{R_{sU,j}^{N}<R_{min}}p_{j,sU}.
\end{align}

Finally, the eventual loss probability is delivered by
\begin{align}\label{eq:Q_final}
Q_s=\frac{\lambda_1+\lambda_2(1-\pi_{sU})}{\lambda}\pi_{sL}+
\frac{\lambda_2\pi_{sU}}{\lambda} Q_{sU}.
\end{align}

\section{Numerical results}\label{sect:num}

In this section, we present our numerical results. Dealing with complex service strategies, we first start assessing the performance of the random access procedure as a function of system parameters. Then, we proceed by analyzing the response of the system metrics including eventual session loss probability and attained rate of NR-U sessions in the unlicensed band. Finally, we compare the proposed offloading strategies. The parameters are provided in Table \ref{table:parameters}. 

\begin{figure*}[!ht]
\centering
\subfigure[{Different $\lambda_B$}]{
	\includegraphics[width=0.28\textwidth]{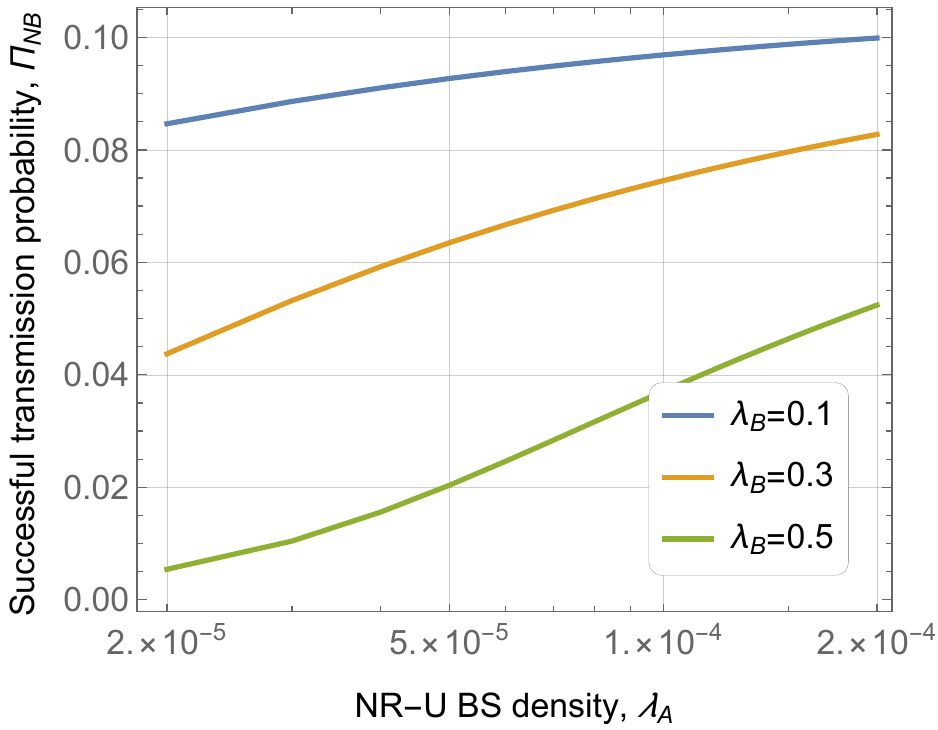}
	\label{fig:suc_1}
}~~~~
\subfigure[{Different contention windows}]{
	\includegraphics[width=0.28\textwidth]{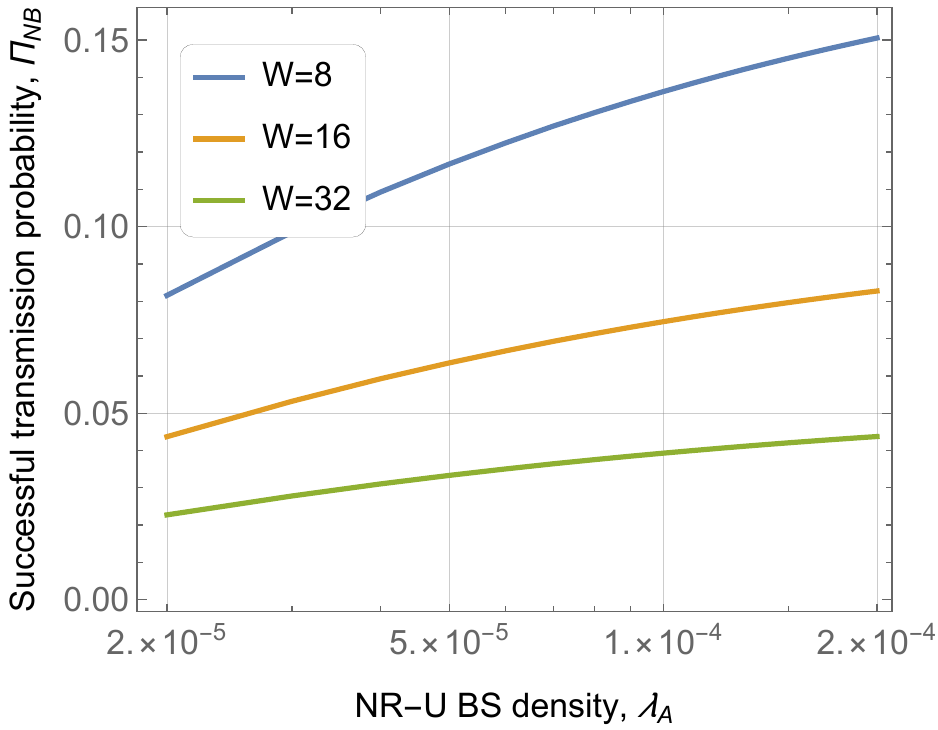}
	\label{fig:suc_2}
}~~~~
\subfigure[{Different session intensity}]{
	\includegraphics[width=0.28\textwidth]{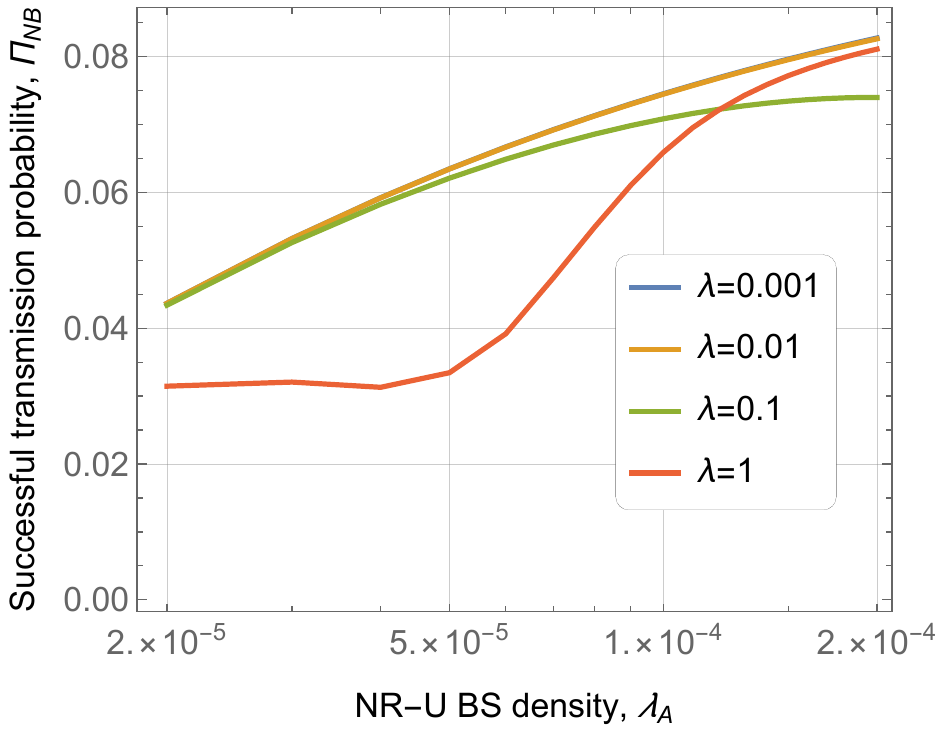}
	\label{fig:suc_3}
}
\caption{Successful NR-U UE transmission probability as a function of system parameters.}
\label{fig:success}
\vspace{-3mm}
\end{figure*}

\begin{figure*}[!t]
\centering
\subfigure[{Different $\lambda_B$}]{
	\includegraphics[width=0.28\textwidth]{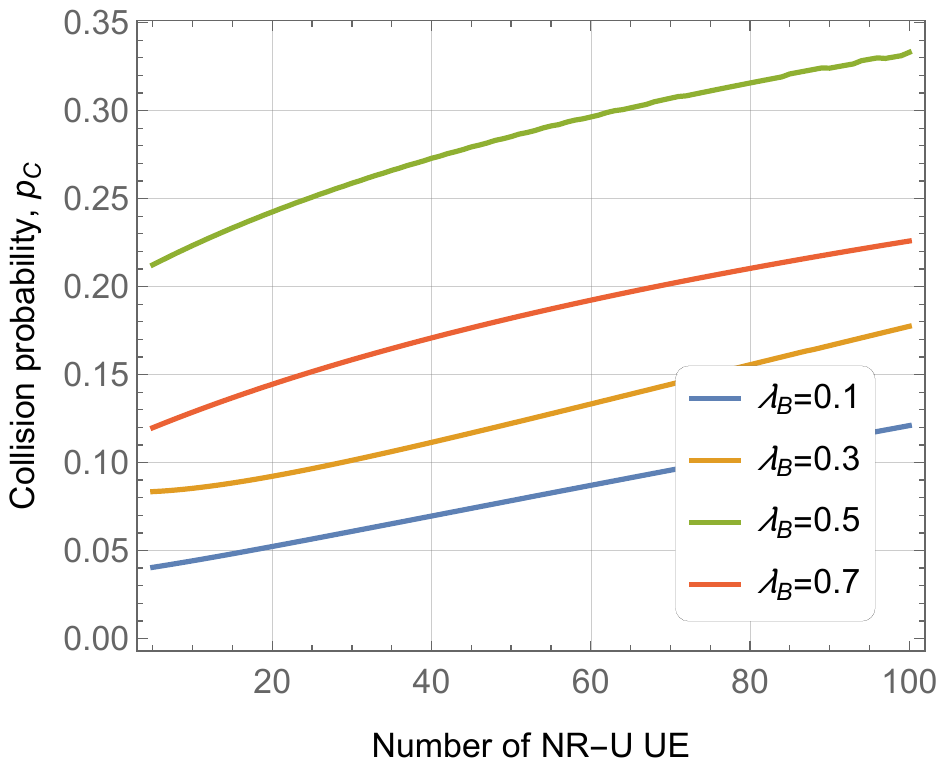}
	\label{fig:coll_1}
}~~~~
\subfigure[{Different contention windows}]{
	\includegraphics[width=0.28\textwidth]{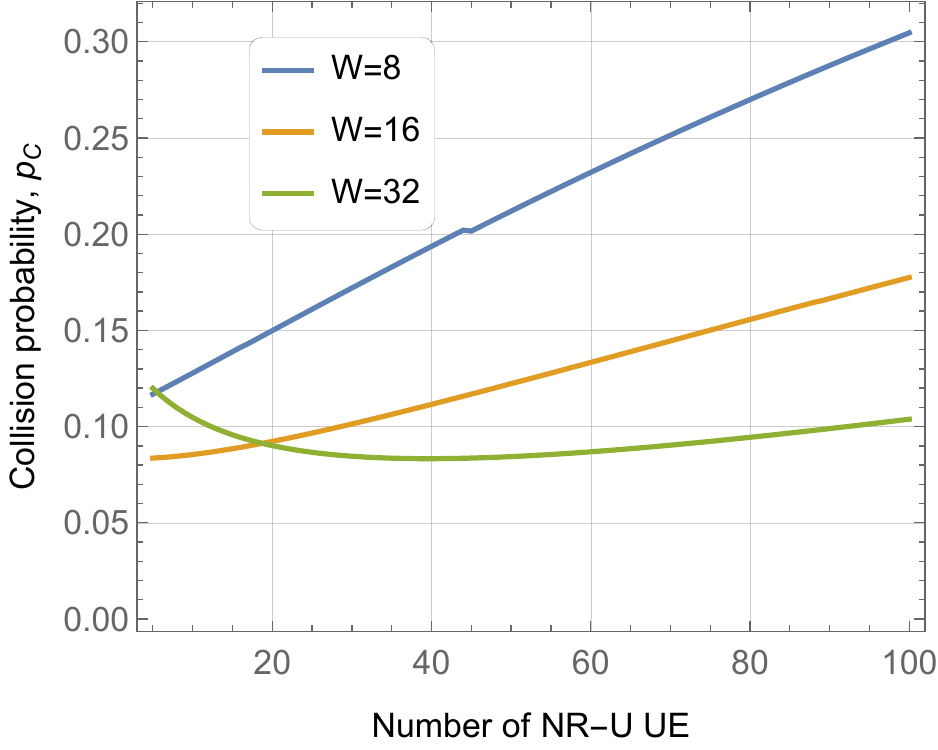}
	\label{fig:coll_2}
}~~~~
\subfigure[{Different number of retransmissions}]{
	\includegraphics[width=0.28\textwidth]{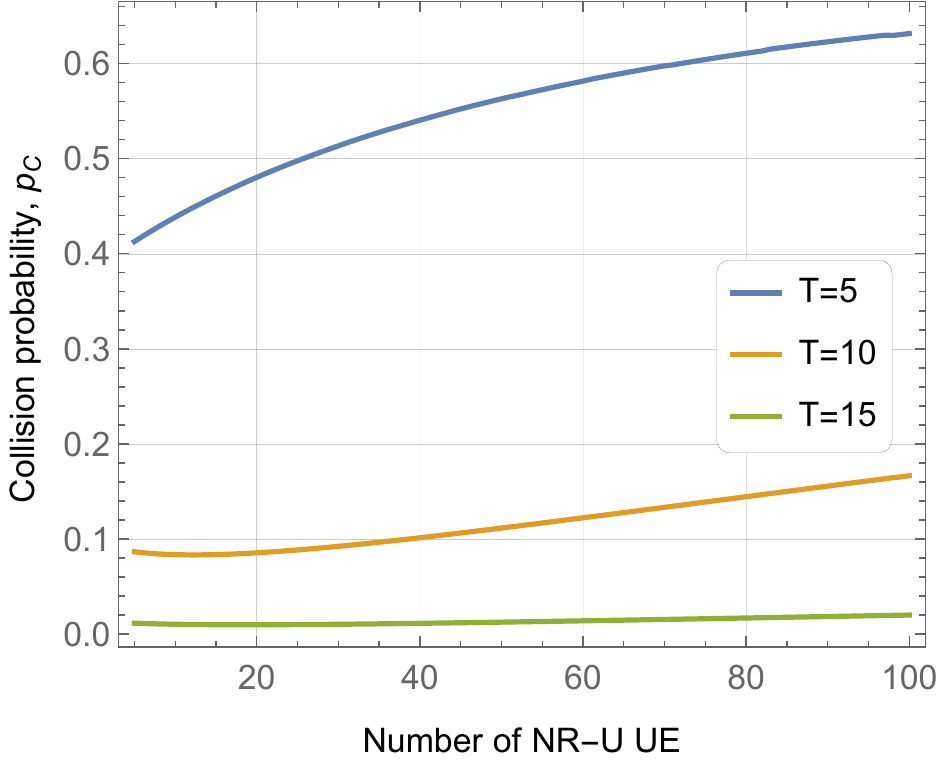}
	\label{fig:coll_3}
}
\caption{NR-U UE collision probability as a function of system parameters.}
\label{fig:collsion}
\vspace{-3mm}
\end{figure*}

\subsection{Random Access Procedure}

We start by analyzing performance of random access mechanisms as a function of system parameters. To this aim, Fig. \ref{fig:collsion} demonstrates collision probability as a function of the number of active NR-U UEs, density of blockers, $\lambda_B$, initial contention window, $W$, and the number of retransmission attempts, $T$. 

\begin{table}[!b]
\vspace{-3mm}
\renewcommand{\arraystretch}{1}
\caption{Default system parameters.}
\label{table:parameters}
\centering
\begin{tabular}{p{0.60\columnwidth}|p{0.25\columnwidth}}
\hline
\hline
\bfseries Parameter & \bfseries Value \\
\hline
Carrier frequencies of NR/WiGig technoligies & 28/60 GHz\\
\hline
Bandwidths of NR/WiGig technologies & 400/2160 MHz\\
\hline
NR-U BS and WiGig AP heights & 10 m\\
\hline
Radius of blockers & 0.2 m\\
\hline
UE and blockers height & 1.7/1.5 m\\
\hline
WiGig and NR transmit powers & 23/33 dBm\\
\hline
Thermal noise & -174 dBm/Hz\\
\hline
Interference margin & 3 dB\\
\hline
NR MCS outage threshold & -8.97 dB\\
\hline
Density of blockers & 0.3\\
\hline
BS/UE antenna arrays of NR technology & 64x4, 8x4\\
\hline
AP/UE antenna arrays of WiGig technology & 16x4, 8x4\\
\hline
NR-U UE active session probability & 0.1\\
\hline
WiGig UE acctive session probability & 0.1\\
\hline
Initial CW & 16\\
\hline
\end{tabular}
\end{table}


By analyzing the data presented in Fig. \ref{fig:coll_1} one may observe that, logically, the collision probability increases when the number of NR-U UEs grow. However, the dependence of the density of blockers is more complicated. Particularly, the increase in the value of $\lambda_B$ first increases the collision probability. This is explained by the fact that more UEs experience blockage once they win the contention for medium. However, maximizing at around $\lambda=0.5$ further increase in blockers density conversely leads to the decrease in the collision probability. This effect is caused by the increase in the value of the current contention window size occurring as a result of unsuccessful transmission attempts. The considered metric also heavily depends on other values of system parameters including the initial contention window and the number of retransmission attempts. This behavior negatively affects the delay performance of the system.


The effect of the initial contention window on the collision probability is demonstrated in Fig. \ref{fig:coll_2}. Here, we again observe behavior that is non-characteristic for lower frequency systems, especially, for larger values of the initial contention window. For small values of the contention window (e.g., $8$ and $16$) and considered blockage probability of $\lambda_B=0.3$ we observe linear increase in the collision probability. However, for $W=32$ the behavior of the considered metric is more complex with clear minimum attained at approximately $40$ UEs. This is explained by the effect of the blockage probability that positively affects the collision probability for small number of active UEs. However, when the number of UEs grows the collision probability start to increase again.


The effect of the retransmission attempts is shown in Fig. \ref{fig:coll_3}. We see the logical decrease in the collision probability when the number of retransmissions increases. Still, for a small number of active UEs, we notice the effect of blockers that decreases the collision probability. However, starting from already 10 UEs both curves show consistent linear increase.

\subsection{Baseline strategy}


Collision probability presented in Fig. \ref{fig:collsion} is a function of a fixed number of active UEs in the system and thus does not account for system deployment parameters. Thus, we now proceed analyzing performance of the baseline strategy starting with the response of the random access mechanism to the considered system deployment. To this aim, Fig. \ref{fig:success} shows the successful transmission probability as a function of the density of NR-U BSs, density of blockers, $\lambda_B$, initial contention window, $W$, and the session intensity of NR-U UEs, $\lambda$.


By analyzing the data presented in Fig. \ref{fig:suc_1} one may observe, that the increase in the blockers density leads to drastic decrease in the value of the considered metric. This behavior is preserved across the whole range of considered NR-U BS deployment densities. Particularly, by increasing the $\lambda_B$ from $0.1$ to $0.5$ we observe similar decrease in the successful transmission probability for NR-U BS density of $2\times{}10^{-5}$. However, for dense deployments with density of $2\times{}10^{-4}$ the decrease is just by one half. Also, observe that the considered metric increases as the NR-U BS density increases. This behavior is explained by the fact that the increase in the NR-U BS deployment density leads to the smaller coverage areas of each BS and thus smaller mean value of the resources needed to provide a given rate. As a result, fewer UEs are offloaded onto the unlicensed band.


Analyzing the data further, we observe that the results presented in Fig. \ref{fig:suc_2} imply that small value of the contention window leads to better system performance for all the considered NR-U densities. The rationale is that under the considered intensity of sessions, the instantaneous number of active UEs is rather small. To explain this effect, consider Fig. \ref{fig:suc_3} showing the dependence of the successful transmission probability on NR-U BS deployment density for different values of the NR-U session intensity, $\lambda$. Here, as one may observe, higher session intensities logically lead to lower successful transmission probabilities. However, this trend is non-uniform for different values of $\lambda$ and NR-U BS densities and also depends on the amount of resources requested by a session. Particularly, for extremely high session arrival intensity of $\lambda=1$ the successful transmission probability slightly decreases up until the NR-U BS density, where the coverage of unlicensed technology becomes comparable to that one of the licensed one (approximately $5\times{}1-^{-5}$). At the same time, in this range of NR-U BS densities, other values of $\lambda$ demonstrate much higher and increasing successful transmission probability. The rationale is that for the latter loads (i) the number of active sessions that can be offloaded onto unlicensed band gradually increases due to the coverage of NR-U BS becomes smaller while (ii) the session drop rerouting probability becoming smaller. For high intensity of sessions, the licensed and unlicensed bands are both overloaded, and thus the intensity of sessions offloaded onto unlicensed technology is virtually constant. When the density of NR-U BS increases further approaching that of the unlicensed technology, the amount of requested resources by a single session at NR-U BS decreases, leading to the proportional decrease in the session rerouting probability and in the successful transmission probability.



\begin{figure}[!t]
\vspace{-0mm}
\centering
\subfigure[{}]{
	\includegraphics[width=0.315\textwidth]{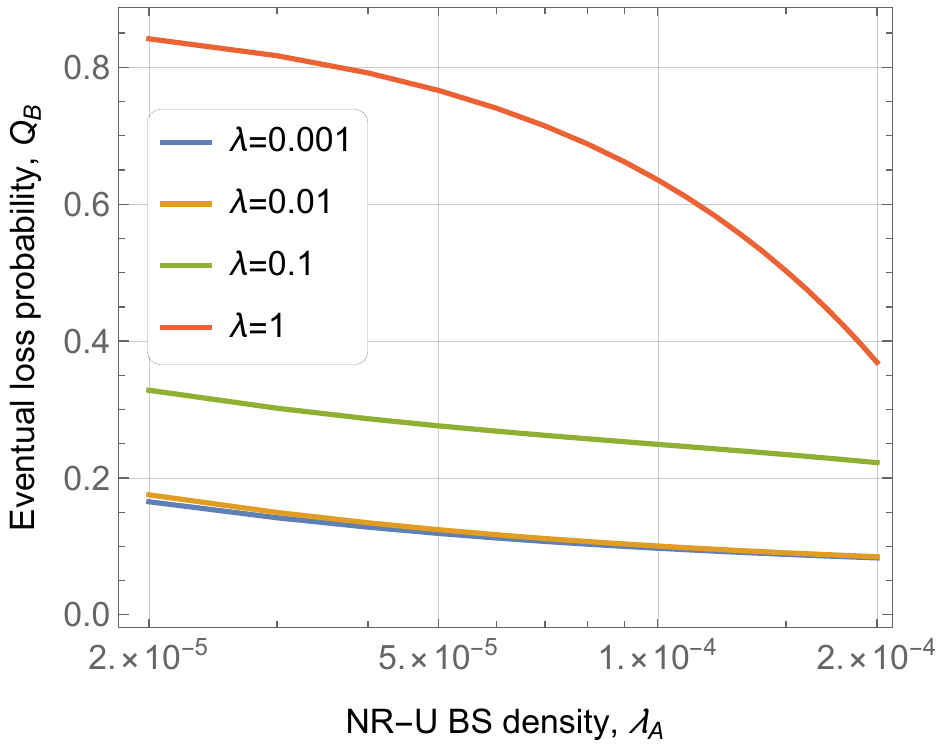}
	\label{fig:loss_1}
}
\subfigure[{}]{
	\includegraphics[width=0.315\textwidth]{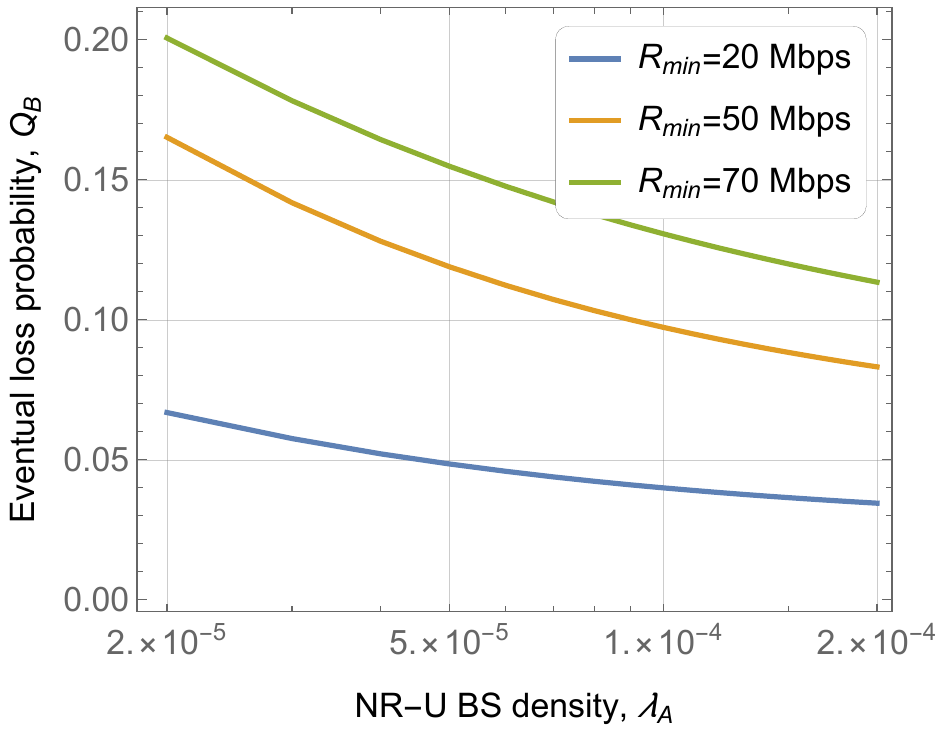}
	\label{fig:loss_2}
}
\caption{Eventual NR-U session loss probability.}
\label{fig:loss}
\vspace{-3mm}
\end{figure}

\begin{figure}[b!]
\vspace{-3mm}
\centering
\includegraphics[width=0.33\textwidth]{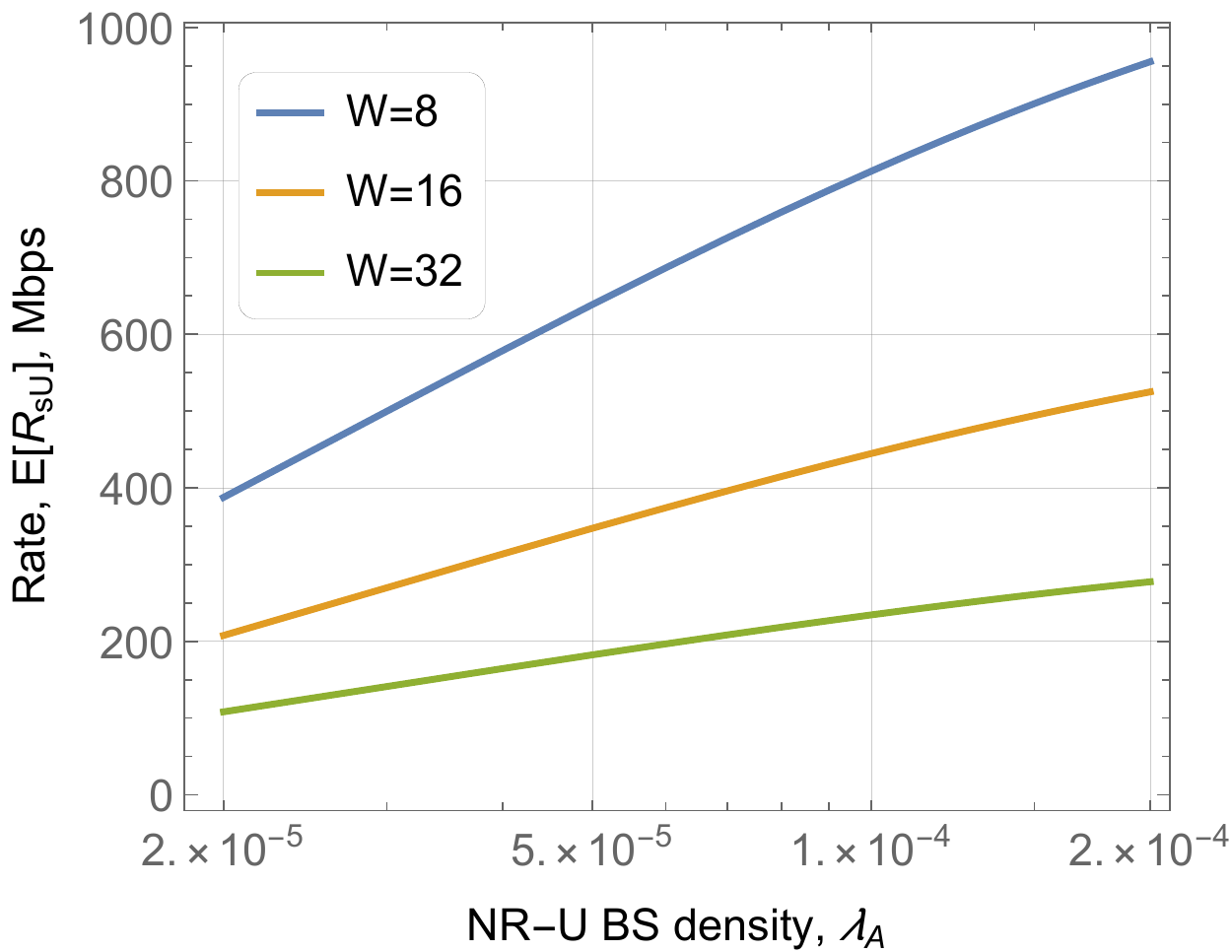}
\caption{Attained NR-U session rate in the unlicensed band.}
\label{fig:unlicensedRate}
\vspace{-0mm}
\end{figure}


Similarly to the successful transmission probability, the eventual session drop probability, illustrated in Fig. \ref{fig:loss} is also a complex function of NR-U BS deployment density, session arrival intensity and the amount of requested resources by a session. Here, we see that Fig. \ref{fig:loss_1} supports our conclusions stated for successful transmission probability in Fig. \ref{fig:suc_3}. More specifically, for all the considered session arrival intensities the system is in fact in the overloaded conditions for low density of NR-U BSs. When the latter parameter increases, the mean amount of resources drastically decreases overweighting the effect of the reduced coverage area of NR-U BS (and thus session arrival intensity that is kept constant). Similar trends are observed for other values of $\lambda$.

The eventual drop probability is represented as a function of the minimum requested session rate, $R_{\min}$ in Fig. \ref{fig:loss_2}. Here, we may observe that the considered metric logically increases when the session requested rate increases. Note that this effect holds for both licensed and unlicensed bands simultaneously as smaller amount of requested resources decreases the loss probability at the licensed band and also impose milder requirements to the amount of required resources in the unlicensed band.


Complementing the analysis above is the attained session rate by offloaded session presented in Fig. \ref{fig:unlicensedRate} as a function of the contention window for the baseline strategy. Recall that this is in fact an intermediate metric that partially reflects whether the session offloaded into unlicensed band is eventually dropped or not. Here, we see that further control (by varying the initial contention window) can be implemented in the unlicensed band to improve the session performance. On top of this, the attained rates also confirm that under $\lambda=0.001$ the amount of offloaded sessions to the unlicensed band is negligible warranting the behavior of successful transmission probabilities in Fig. \ref{fig:success} and the associated attained rates.

\subsection{Comparison of Considered Strategies}

\begin{figure}[!t]
\vspace{-0mm}
\centering
\subfigure[{Session rate $50$ Mbps}]{
	\includegraphics[width=0.31\textwidth]{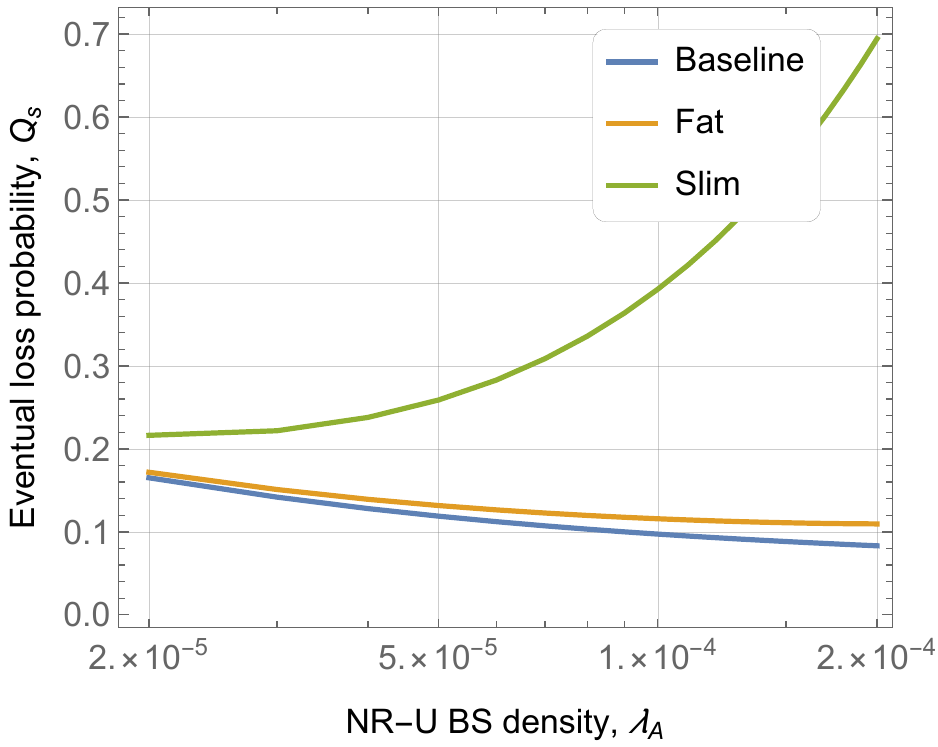}
	\label{fig:str_1}
}
\subfigure[{Session rate $100$ Mbps}]{
	\includegraphics[width=0.31\textwidth]{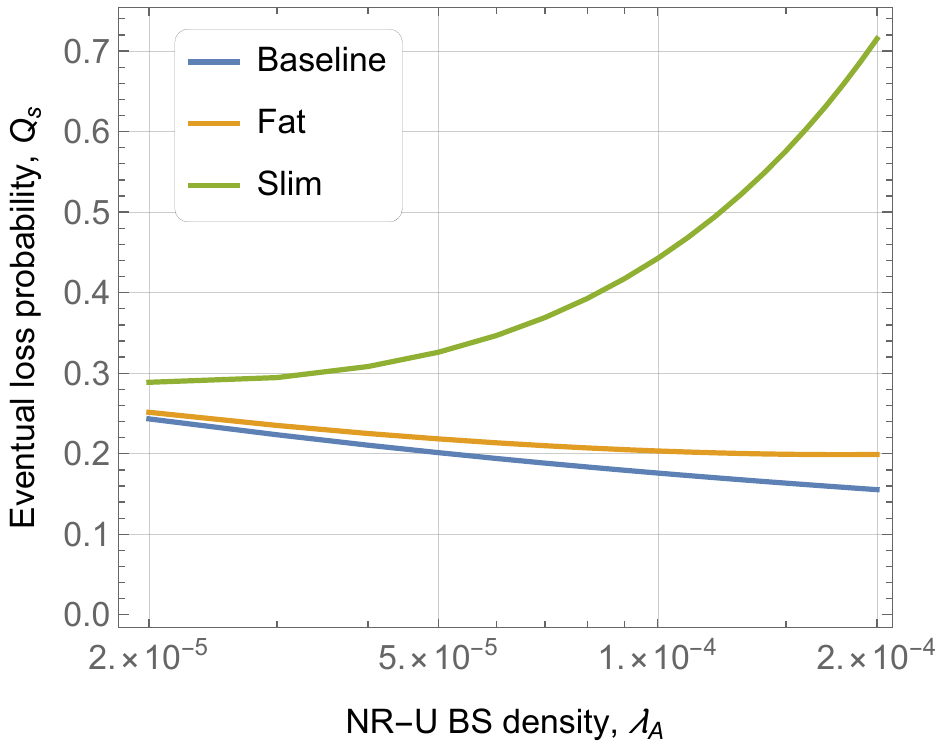}
	\label{fig:str_2}
}
\caption{Eventual session loss probability for considered strategies.}
\label{fig:str}
\vspace{-3mm}
\end{figure}


\begin{figure*}[!ht]
\centering
\subfigure[{Same contention windows}]{
	\includegraphics[width=0.28\textwidth]{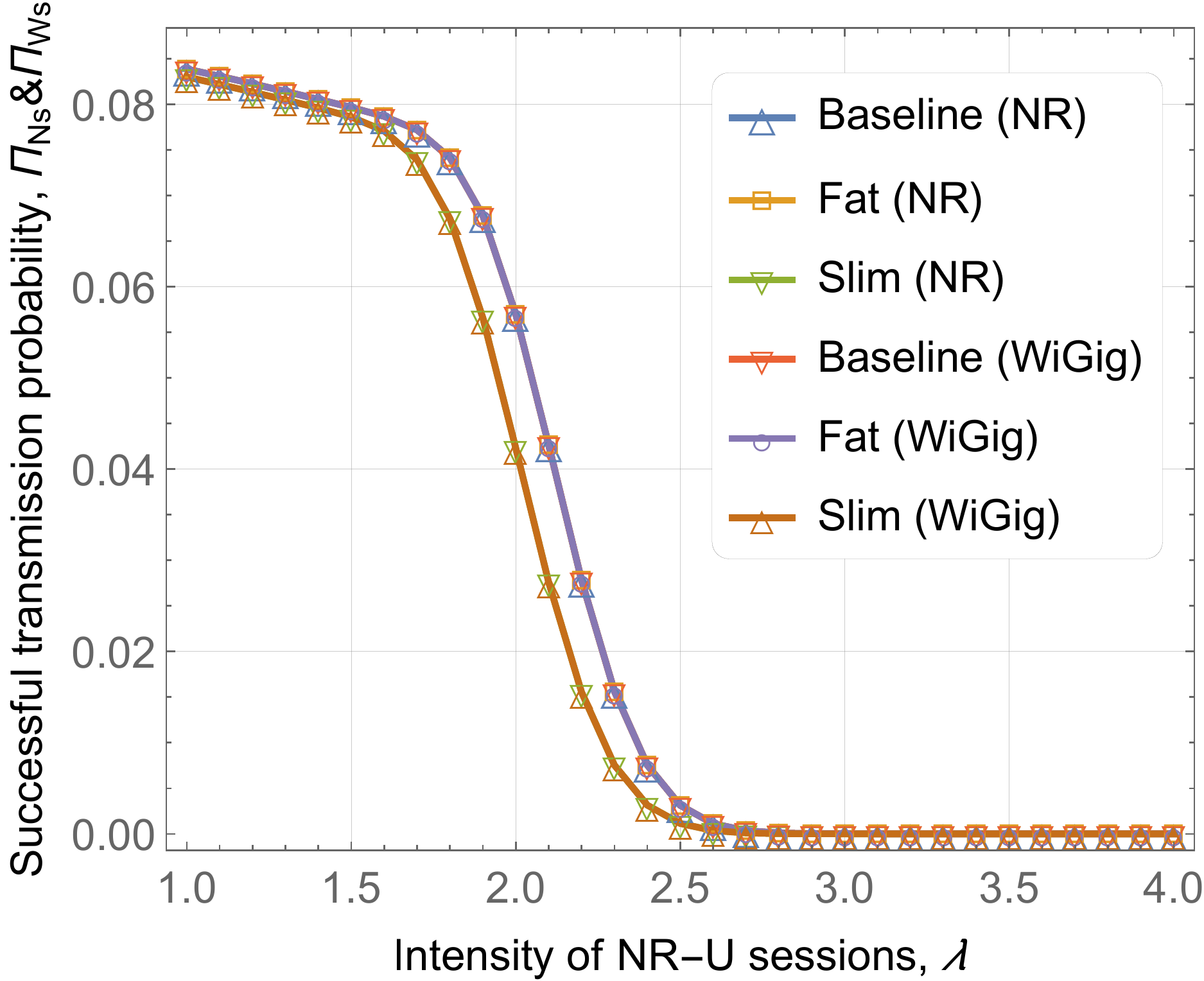}
	\label{fig:ans_2}
}~~~~
\subfigure[{Different contention windows}]{
	\includegraphics[width=0.28\textwidth]{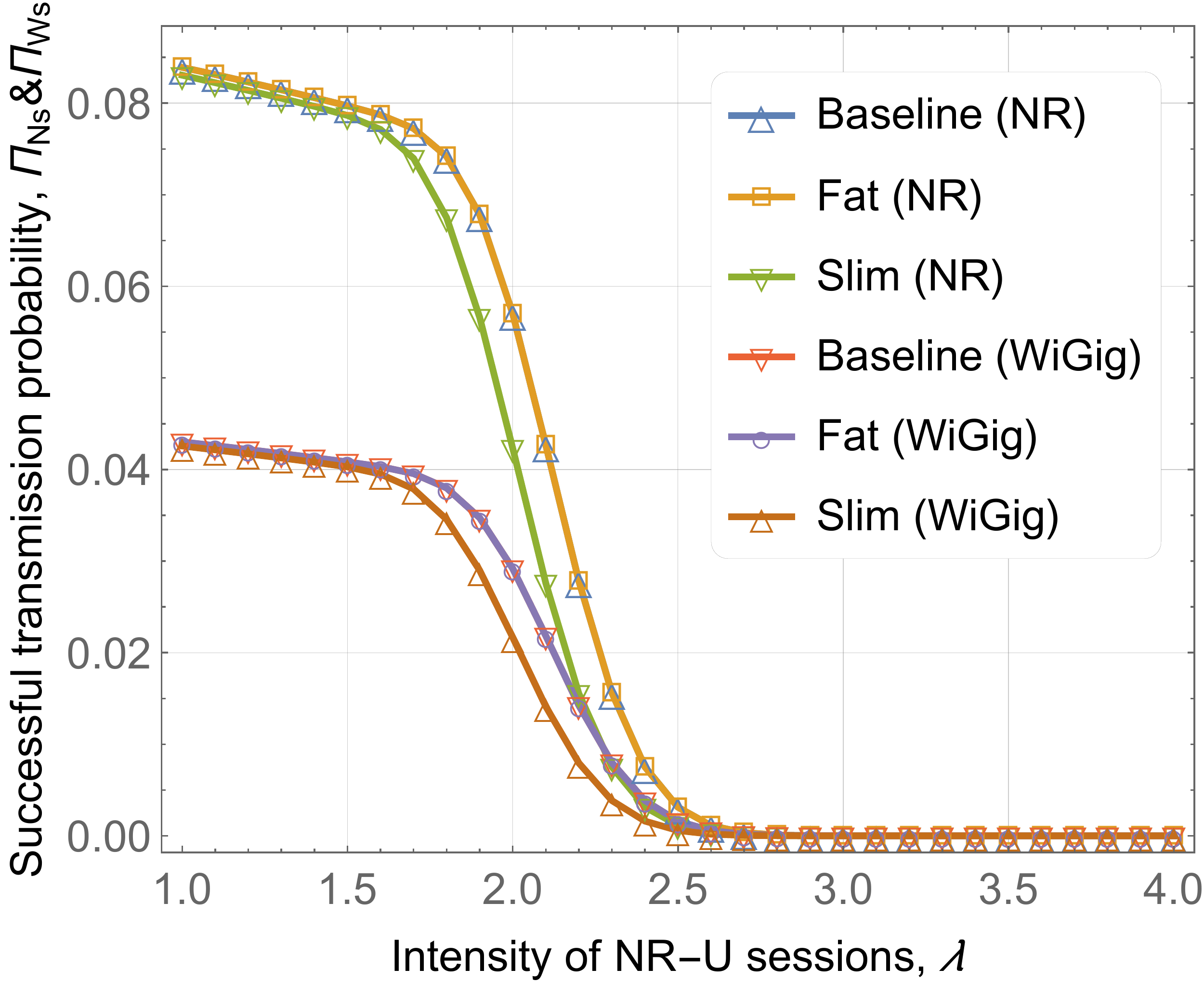}
	\label{fig:ans_3}
}~~~~
\subfigure[{Rate for the different contention windows}]{
	\includegraphics[width=0.28\textwidth]{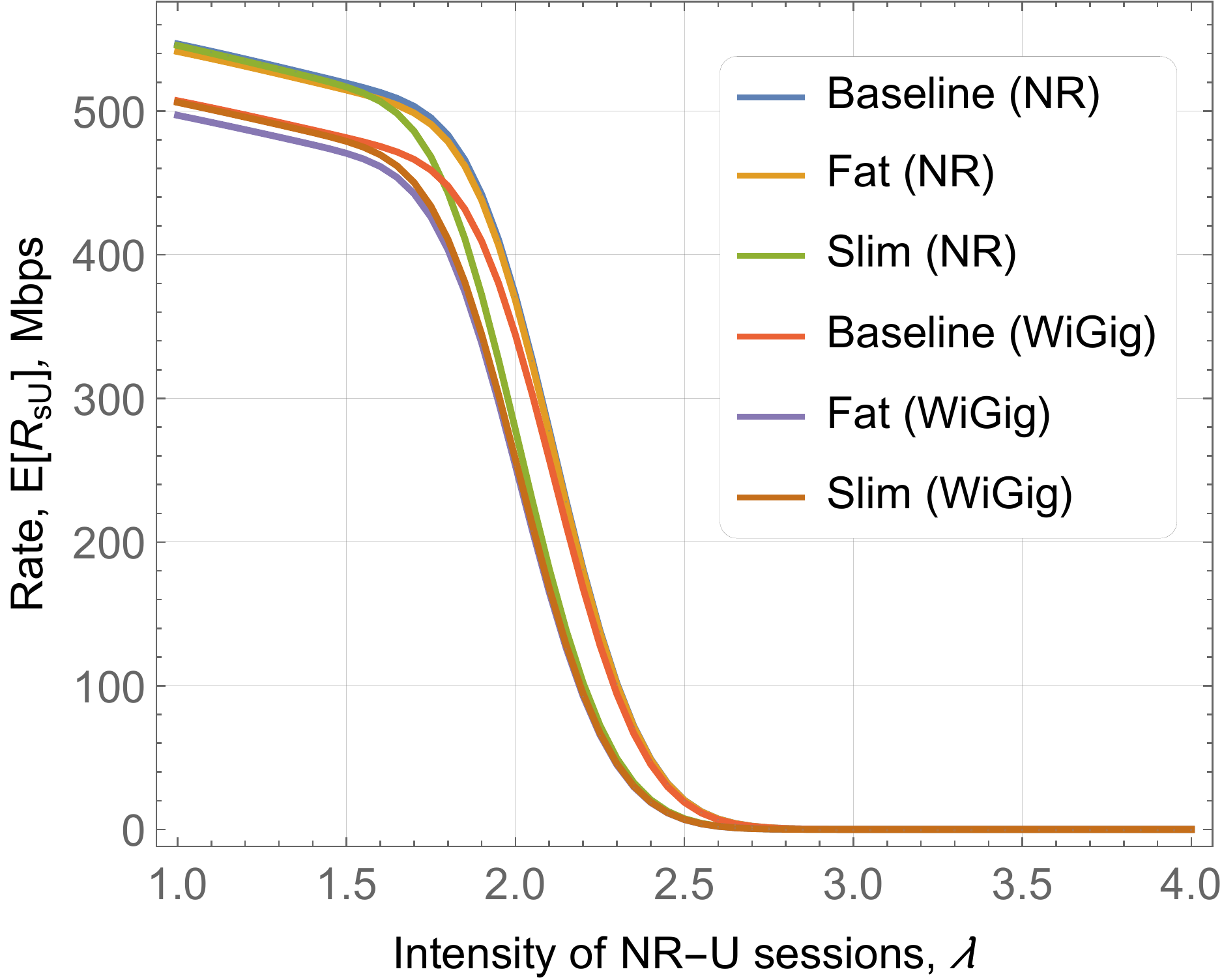}
	\label{fig:ans_1}
}
\caption{Attained rate in the unlicensed band of NR-U and WiGig UEs and successful NR-U UE transmission probability as a function of system parameters.}
\label{fig:answ}
\vspace{-3mm}
\end{figure*}

Having studied the random access mechanism under blockage impairments and baseline scheme we are now in position to proceed with comparison of the considered offloading strategies. To this aim, below we consider three variants of the strategies: (i) baseline, where a session is offloaded when no resources are available at NR part of NR-U BS, (ii) slim strategy with $R_{S}$ set to the mean resource requirements of a session and (iii) fat strategy with $R_{S}$ set to the mean resource requirements of a session.

Fig. \ref{fig:str} illustrates the eventual session loss probability for all considered strategies and two different session resource requirements, $50$ Mbps and $100$ Mbps. By analyzing the presented results one may observe that the baseline strategy, where a session is offloaded onto unlicensed band when no resources for its service are available in the licensed one, is associated with the minimal values of the eventual session drop probabilities. At the same time, offloading heavy sessions to the unlicensed band leads to very similar performance while "slim" strategy, where "lighter" sessions are offloaded onto unlicensed band is characterized by drastically worse performance. The latter can be explained by the fact, that random access mechanisms under the same offered load are known to utilize available resources better when there are few "heavy" sessions in the system rather than a plethora of "lighter" sessions. This is also confirmed by the increase in the session loss probability for the slim strategy when the density of the NR-U BS deployment increases. One may further observe, that the response of the system is not affected qualitatively by the requested session rates. Quantitatively, however, the increase of the requested rate logically leads to the associated increase in the eventual session loss probability.

\subsection{Fairness Between WiGig and NR-U UEs}

We now show the fairness between NR-U and WiGig UEs in the unlicensed band. To this aim, Fig. \ref{fig:answ} shows the successful transmission probability for NR-U $\Pi_N$ and WiGig UEs $\Pi_W$ and attained rate in the unlicensed band of NR-U $E[R_{sU}^N]$ and WiGig UEs  $E[R_{sU}^W]$ for different rerouting strategies and initial contention windows. As one may observe, the successful transmission probability for NR-U $\Pi_N$ and WiGig UEs $\Pi_W$ and attained rate in the unlicensed band of NR-U $E[R_{sU}^N]$ and WiGig UEs $E[R_{sU}^W]$ logically decrease as the intensity of NR-U sessions increases $\lambda$. The drop is slow at first, as most of the collisions are resolved, and then there is a sharp drop for both types of UEs caused by overloaded conditions.

As one may observe, we see that the successful transmission probabilities for both types of devices coincide at 16, see Fig. \ref{fig:ans_2}, implying that the fairness is retained with respect to this metric. However, this naturally leads to different attained rates in the unlicensed band due to the difference in the MCS utilized by NR and WiGig. If one requires fairness in terms of the data rates, one may adjust the CW size. To this aim. Fig. \ref{fig:ans_3} shows successful transmission probability, where the initial CW size of NR-U UEs is two times higher. Here, we see that this metric becomes worse for NR-U UEs translating to equal rates of both types of UEs as shown in Fig. \ref{fig:ans_1}.


\section{Conclusions}\label{sect:concl}

In this paper, inspired by the potential use of NR-U technology for smoothing short-term traffic variations in the licensed band, we have developed an analytical framework that captures mmWave-specific propagation, service process in the licensed and unlicensed band, and LBT-based coexistence scheme of NR-U technology. To this aim, we merged the tools of stochastic geometry, renewal theory, and queuing systems with random resource requirements. We then proceeded to study different offloading strategies using the eventual NR-U session loss as the main metric of interest.

Empowered with the developed framework, we proceeded analyzing different offloading strategies. Our results demonstrate non-trivial behavior of the collision probability as compared to lower frequency systems with blockage having non-linear impact: the increase in the density of blockers first increases the collision probability as a result of more UEs experiencing collision upon winning of contention and then decreases due to increase in the current contention window at UEs. This effect is characteristic for small to moderate number of active UEs in the system (up to $10-20$) and heavily depends on the choice of other system parameters the initial contention window and the number of retransmission attempts.

Further, by comparing the offloading strategies we revealed that the baseline strategy, where a session is offloaded onto unlicensed band only when there are no resources available in the licensed one, leads to the best performance. However, the "fat" strategy that initially offloads heavier sessions onto unlicensed band, is only $2-5\%$ worse in terms of the eventual session loss probability. This observation may allow to design efficient traffic offloading strategies at UE side. By utilizing the proposed model one may estimate the deployment density of the collocated NR-U/WiGig BSs resulting in a prescribed session loss probability as well as fine-tune the performance of the unlicensed band to further decrease the latter probability.




\bibliographystyle{IEEEtran}
\bibliography{main3}

\end{document}